\documentclass[fleqn,usenatbib]{mnras}

\usepackage[T1]{fontenc}

\DeclareRobustCommand{\VAN}[3]{#2}
\let\VANthebibliography\thebibliography
\def\thebibliography{\DeclareRobustCommand{\VAN}[3]{##3}\VANthebibliography}

\usepackage{graphicx}	
\usepackage{amsmath}	
\usepackage{amssymb}	

\usepackage{xcolor}

\usepackage{multirow}
\usepackage{booktabs}

\usepackage{newtxtext,newtxmath}

\usepackage{algorithm} 
\usepackage{algpseudocode} 

\usepackage{numprint}

\usepackage{subcaption}

\let\vec\mathbf

\newcommand{\Mpch}{h^{-1}\mathrm{Mpc}}

\newcommand{\av}[1]{\left\langle{#1}\right\rangle}

\newcommand{\vk}{\vec k}

\newcommand{\vs}{\vec s}

\newcommand{\vr}{\vec r}

\newcommand{\ft}[1]{\mathcal{F}\left[{#1}\right]}
\newcommand{\ift}[1]{\mathcal{F}^{-1}\left[{#1}\right]}

\newcommand{\hR}{\hat{\vec R}}
\newcommand{\vR}{\vec R}
\newcommand{\Ng}{N_\mathrm{g}}
\newcommand{\hr}{\hat{\vec r}}

\newcommand{\tjo}[3]{\begin{pmatrix} {#1} & {#2} & {#3}\\ 0 & 0 & 0\end{pmatrix}}
\newcommand{\tj}[6]{\begin{pmatrix} {#1} & {#2} & {#3}\\ {#4} & {#5} & {#6}\end{pmatrix}}

\newcommand{\pluseq}{\mathrel{+}=}

\def\beq{\begin{eqnarray}}
\def\eeq{\end{eqnarray}}

\definecolor{darkgreen}{RGB}{0,120,0}

\newcommand{\resub}[1]{#1}

\renewcommand{\L}{\Lambda}
\renewcommand{\P}{\mathcal{P}}

\title[Estimating $N$-point Correlation Functions with \textsc{encore}]{\textsc{encore}: \resub{An $\mathcal{O}(N_g^2)$ Estimator for Galaxy $N$-Point Correlation Functions}}

\author[Philcox \textit{et al.}]{Oliver H.\,E. Philcox$^{1,2}$\thanks{E-mail: \href{mailto:ohep2@cantab.ac.uk}{ohep2@cantab.ac.uk} (OP)},
Zachary Slepian$^{3,4}$, Jiamin Hou$^{3}$, Craig Warner$^{3}$,
\newauthor
Robert N. Cahn$^{4}$ and Daniel J. Eisenstein$^{5}$
\\
$^{1}$Department of Astrophysical Sciences, Princeton University, Princeton, NJ 08540, USA\\
$^{2}$School of Natural Sciences, Institute for Advanced Study, 1 Einstein Drive, Princeton, NJ 08540, USA\\
$^{3}$Department of Astronomy, University of Florida, 211 Bryant Space Science Center, Gainesville, FL 32611, USA\\
$^{4}$Physics Division, Lawrence Berkeley National Laboratory, 1 Cyclotron Road, Berkeley, CA 94709, USA\\
$^{5}$Center for Astrophysics | Harvard \& Smithsonian, 60 Garden St., Cambridge, MA 02138, USA\\
}

\pubyear{2021}

\begin{document}
\label{firstpage}
\pagerange{\pageref{firstpage}--\pageref{lastpage}}
\maketitle


\begin{abstract}
We present a new algorithm for efficiently computing the $N$-point correlation functions (NPCFs) of a 3D density field for arbitrary $N$. This can be applied both to a discrete \resub{spectroscopic} galaxy survey and a continuous field. By expanding the statistics in a separable basis of isotropic functions built from spherical harmonics, the NPCFs can be estimated by counting pairs of particles in space, leading to an algorithm with complexity $\mathcal{O}(\Ng^2)$ for $\Ng$ particles, or $\mathcal{O}\left(N_\mathrm{FFT}\log N_\mathrm{FFT}\right)$ when using a Fast Fourier Transform with $N_\mathrm{FFT}$ grid-points. In practice, the rate-limiting step for $N>3$ will often be the summation of the histogrammed spherical harmonic coefficients, particularly if the number of \resub{radial and angular} bins is large. In this case, the algorithm scales linearly with $\Ng$. The approach is implemented in the \textsc{encore} code, which can compute the 3PCF, 4PCF, 5PCF, and 6PCF of a BOSS-like galaxy survey in $\sim$ $100$ CPU-hours, including the corrections necessary for non-uniform survey geometries. We discuss the implementation in depth, along with its GPU acceleration, and provide practical demonstration on realistic galaxy catalogs. Our approach can be straightforwardly applied to current and future datasets to unlock the potential of constraining cosmology from the higher-point functions.
\end{abstract}

\begin{keywords}
methods:  statistical,  numerical  --  Cosmology:  large-scale  structure of Universe, theory -- galaxies: statistics
\end{keywords}



\section{Introduction}\label{sec: intro}
Amongst the most powerful tools in the survey analyst's workshop are the $N$-point correlation functions (NPCFs), or their Fourier-space counterparts, the polyspectra. These encode the statistical properties of the galaxy overdensity field at sets of $N$ positions, and may be compared to data to give constraints on properties such as the Universe's expansion rate and composition. Most inflationary theories predict density fluctuations in the early Universe to follow Gaussian statistics; in this case all the information is contained within the two-point correlation function (2PCF), or, equivalently, the power spectrum. 
For a homogeneous and isotropic Universe, both are simple functions of one variable, and have been the subject of almost all 
galaxy survey analyses to date \citep[e.g.,][]{2017MNRAS.470.2617A,2017MNRAS.466.2242B,2021PhRvD.103h3533A,2020JCAP...05..042I, 2020JCAP...05..032P,2020JCAP...05..005D,2020JCAP...06..001C}. 

The late-time Universe is far from Gaussian. Statistics beyond the 2PCF have been long discussed in the literature \citep[e.g.,][]{1975ApJ...196....1P,peebles78,2001ASPC..252..201P,2000ApJ...544..597S,2002PhR...367....1B,2005astro.ph..2088C}, and, as shown in \citet{2015PhRvD..92l3522S}, 
are of importance since the bulk motion of matter in structure formation causes a cascade of information from the 2PCF to higher-order statistics, such as the three- and four-point functions (3PCF and 4PCF).
Recent work by \citet{2021JCAP...07..008G} demonstrated the power of such correlators; for the matter field, forecasted constraints on key cosmological parameters were shown to tighten significantly when the bispectrum and trispectrum (the Fourier-space counterparts to the 3PCF and 4PCF) were included in the analysis. A number of works have demonstrated similar effects for the galaxy density field, including improved constraints on $\L$CDM parameters \citep{2017MNRAS.467..928G,2021JCAP...03..021A}, neutrino masses \citep{2019JCAP...11..034C,2020arXiv201100899K,2021JCAP...04..029H}, modified gravity \citep{2020arXiv201105771A}, and primordial non-Gaussianity \citep{2018MNRAS.478.1341K,2021JCAP...05..015M}. \citet{2021MNRAS.505..628S} provides another notable example, demonstrating that, when higher-order correlators are included, parameters such as the Alcock-Paczynski dilation statistic $\alpha$ can be constrained more strongly from late-time measurements than from the initial matter power spectrum. This occurs since such parameters (which are a key source of information on dark energy) appear only in the forward model rather than in the initial conditions. It remains to be seen whether the purported gains hold in the more realistic scenarios of shot-noise limited galaxy surveys. 

The above discussion suggests that large-scale structure analyses should certainly be including statistics beyond the 2PCF. This is far from trivial however, since the higher-order NPCFs (or polyspectra) are (a) expensive to estimate, (b) difficult to model and (c) of high dimension. Na\"ive NPCF estimation requires counting all $N$-tuplets of galaxies (and random particles) and assigning them to bins; whilst possible for $N=2$, this is prohibitively slow for the higher-point statistics, and scales as $\Ng^N$ for $\Ng$ particles. For the simplest beyond-Gaussian statistic, the isotropic 3PCF, a number of algorithms of varying efficiency have been proposed to circumvent this problem, using techniques such as $kd$-trees, Fourier Transforms, and basis decompositions (e.g., \citealt{1998ApJ...494L..41S,2001misk.conf...71M,2004ASPC..314..249G,2004ApJ...605L..89S,2005NewA...10..569Z,2015MNRAS.454.4142S,march12,2016MNRAS.455L..31S,2017arXiv170900086F,2018ApJ...862..119P,2020MNRAS.492.1214P,2021MNRAS.501.4004P} as well as \citealt{2019ApJS..242...29S,2019AJ....158..116T} 
for the (possibly compressed) 4PCF.) 

In terms of theory models, whilst a number exist for the galaxy 3PCF and bispectrum \citep[e.g.,][]{2000ApJ...544..597S,2003MNRAS.340..580T,2004MNRAS.353..287W,2005MNRAS.361..824G,2008ApJ...672..849M,2017MNRAS.469.2059S,2018JCAP...04..030S,2021JCAP...05..035U}, 
many of which are semi-analytic, much work remains to be done. For the matter field, perturbative approaches have been computed up to one-loop order \citep{2015JCAP...05..007B,2015JCAP...10..039A,2016JCAP...06..052B},  
but this has not yet been extended to galaxy surveys, hampered by the necessity to include a full set of quartic (quintic) galaxy bias parameters for the 3PCF (4PCF) model. 

Finally, even when the statistics are measured and modeled, extracting information from them is non-trivial due to their high dimension. Accurate analyses are likely to require data compression techniques \citep[e.g.,][]{2000ApJ...544..597S,2000MNRAS.317..965H,2011MNRAS.412.1993M,2016ApJ...827...26B,2017MNRAS.466L..83J,2018MNRAS.476L..60A,2018MNRAS.476.4045G,2018MNRAS.474.2109S,2019MNRAS.484L..29G,2021PhRvD.103d3508P} as well as analytic covariance matrices (e.g., \citealt{2015MNRAS.454.4142S,2018MNRAS.478.1468S,2019MNRAS.490.5931P,2020MNRAS.497.1684S,2021JCAP...05..035U} for the 3PCF).  
To date, the isotropic 3PCF or bispectrum has been included in a number of analyses \citep{1998ApJ...503...37J,2001ApJ...546..652S,2004ApJ...607..140J,2004PASJ...56..415K,2006MNRAS.368.1507N,2011ApJ...737...97M,2015MNRAS.451..539G,2017MNRAS.465.1757G,2017MNRAS.469.1738S,2018MNRAS.474.2109S,2018MNRAS.478.4500P}, but few use any higher-order statistics (though see \citealt{2019ApJS..242...29S}, \citealt{peebles78} and the review of \citealt{2001ASPC..252..201P} for some notable exceptions, as well as \citet{2019PhRvD.100l3511M} for the CMB).

The present work tackles the following problem: how can we efficiently measure the isotropic NPCFs for $N>3$? To do this, our starting point is the isotropic 3PCF algorithm of \citet{2015MNRAS.454.4142S}. By expanding the correlator's angular dependence into a Legendre polynomial basis and further factorizing these into spherical harmonics, 
the former work was able to estimate the 3PCF of $\Ng$ 
particles \resub{with $\mathcal{O}(\Ng^2)$ complexity}; much faster than a na\"ive triple count, which \resub{has a complexity of $\mathcal{O}(\Ng^3)$}. This has facilitated a number of analyses, including detection of the 3PCF baryon acoustic oscillation (BAO) feature and a constraint on the baryon-dark matter relative velocity \citep{2017MNRAS.468.1070S,2017MNRAS.469.1738S,2018MNRAS.474.2109S}. Here, we consider whether such an approach is possible for higher-point functions. A crucial component is the choice of basis; 
for this we will use the separable isotropic $(N-1)$-point basis functions introduced in \citet{2020arXiv201014418C} \resub{(see also \citealt{2020PhRvR...2c3004M})}, which are a generalization of the polypolar spherical harmonics of zero total angular momentum
\citep{1988qtam.book.....V}. Indeed, the principal conclusion of this work is that an $\mathcal{O}(\Ng^2)$ NPCF estimator is possible for \textit{all} $N$; furthermore, its practical implementation scales linearly with $\Ng$ if $N$ is large. In fact, such an estimator can be constructed in arbitrary dimensions, as long as the underlying space is homogeneous and isotropic; generalization to such cases is discussed in our companion work \citet{npcf_generalized}. Below,  
we will present a detection of the galaxy 4PCF and 5PCF by applying our algorithm, implemented in the \textsc{encore} code, to mock catalogs. It is important to note that, for $N>3$, only part of the information in the NPCF is independent from that in lower-point functions; there is also a `disconnected' term that does not add further information.

A key assumption underlying the above discussion is that of isotropy. Arising from the projection of galaxies' peculiar motions onto our line of sight, redshift-space distortions (RSD; \citealt{1987MNRAS.227....1K,1998ASSL..231..185H,1996ApJ...462...25Z}) break this ideal, introducing additional dependence on the line-of-sight angle. However, this effect also adds cosmological information in the form of the growth-rate $f(z)$ and allows us to break degeneracies, for example those between linear galaxy bias and the primordial amplitude $A_s$ in the 2PCF \citep[e.g.,][]{2013PhR...530...87W}. Inclusion of such effects has become commonplace in any 2PCF analysis, but has been limited for higher-point functions, though \citet{2020JCAP...06..041G,2021MNRAS.501.2862S} show anisotropy to contain useful information. 
In general, whilst there are a handful of estimators dedicated to the measurement of the anisotropic 3PCF or bispectrum 
\citep[e.g.,][]{2007arXiv0709.1967G,2015PhRvD..92h3532S,2018MNRAS.478.1468S,2019MNRAS.484..364S,2020arXiv201103503G}, they have been scarcely used. Whilst we consider only the isotropic NPCFs in this work, it is expected that the methodology can be extended further, utilizing basis functions with a non-zero total angular momentum; this is discussed in \citet{npcf_generalized}. 

\vskip 4pt
The remainder of this paper is structured as follows. We begin in \S\ref{sec: basis} by introducing the basis functions underlying our algorithm (following \citealt{2020arXiv201014418C}), before presenting the NPCF estimators in \S\ref{sec: estimator}, including discussion of the split into `connected' and `disconnected' components (\S\ref{subsec: estimator-def}), and the necessary corrections for non-uniform survey geometry (\S\ref{subsec: edge-correction}). \S\ref{sec: implementation} contains a detailed discussion of the \textsc{encore} code, including its CPU and GPU implementations, before we present results in \S\ref{sec: results}, both concerning the algorithm scalings and 
application to mock data. We conclude with a summary in \S\ref{sec: summary}, before presenting several useful mathematical results in Appendices \ref{appen: generalized-gaunt}\,\&\,\ref{appen: spin-sum}. 

\section{Basis Functions}\label{sec: basis}
To begin, we present the isotropic basis functions underlying our NPCF estimators and outline some of their key properties.
Full details of these functions \resub{(which are similar to those of \citealt{2020PhRvR...2c3004M})} are presented in  \citet{2020arXiv201014418C}. Generalization anisotropic functions, as well as homogeneous and isotropic spaces of arbitrary dimension is presented in \citet{npcf_generalized}.

\subsection{Definition}\label{subsec: basis-special}
Consider a function, $f$, depending on $(N-1)$ spatial coordinates in 3D Euclidean space, $\vr_1, \ldots, \vr_{N-1}$.\footnote{Foreshadowing the application to $N$-point functions, we consider only basis functions involving $(N-1)$ coordinates in this section. Statistical homogeneity dictates that the NPCFs are independent of the $N$-th spatial position.} If the function is invariant under simultaneous rotations of all $(N-1)$ coordinates, \textit{i.e.} $f(R\vr_1,\ldots,R\vr_{N-1}) = f(\vr_1,\ldots,\vr_{N-1})$ for any rotation matrix $R$, $f$ can be expanded in a basis of isotropic functions as
\beq\label{eq: isotropic-basis-expansion}
    f(\vr_1,\ldots,\vr_{N-1}) = \sum_{\L} f_{\L}(r_1,\ldots,r_{N-1})\P_{\L}(\hr_1,\ldots,\hr_{N-1}),
\eeq
where the set $\L$ indexes the 
basis function in question; the precise form of this will be given below. To fulfill the requirements of isotropy, the functions must satisfy
\beq\label{eq: basis-rotation}
    \P_\L(R\hR) = \P_\L(\hR)
\eeq
for all rotations $R$, introducing the nomenclature $\vR = \{\vr_1,\ldots,\vr_{N-1}\}$, $\hR = \{\hr_1,\ldots,\hr_{N-1}\}$.

We now consider the functional form of such a basis, following \citet{2020arXiv201014418C} and \citet{npcf_generalized}. Firstly, we note that any scalar function of $(N-1)$ variables in 3D Euclidean space can be expanded in terms of $(N-1)$ spherical harmonics, regardless of isotropy:
\beq\label{eq: generalized-Ylm-expansion}
    g(\vr_1,\ldots,\vr_{N-1}) = \sum_{\ell_1\ldots \ell_{N-1}}\sum_{m_1\ldots m_{N-1}} g_{\ell_1\ldots\ell_{N-1}}^{m_1\ldots m_{N-1}}(r_1,\ldots,r_{N-1}) Y_{\ell_1m_1}(\hr_1)\ldots Y_{\ell_{N-1}m_{N-1}}(\hr_{N-1}),
\eeq
where $Y_{\ell m}(\hr)$ is a spherical harmonic (in the Condon-Shotley phase convention), $\ell$ and $m$ are the orbital and magnetic angular momentum quantum numbers, and we sum over all $\ell_i\geq 0$ and $|m_i|\leq \ell_i$. If we further assume $g$ to be isotropic, relation \eqref{eq: basis-rotation} places constraints on the allowed values of $m_i$; in particular, this enforces $m_1+\ldots+m_{N-1}=0$. 
To obtain an isotropic basis from \eqref{eq: generalized-Ylm-expansion}, we must consider the possible ways in which to combine products of $(N-1)$ spherical harmonics to obtain a rotationally invariant state. As shown in \citet{2020arXiv201014418C}, such an approach leads to the set of basis functions
\beq\label{eq: basis-def}
    \boxed{\P_\L(\hR) = \sum_{m_1\ldots m_{N-1}} C^\L_M\,Y_{\ell_1m_1}(\hr_1)\ldots Y_{\ell_{N-1}m_{N-1}}(\hr_{N-1}),}
\eeq
where $C^\L_M$ is a set of coupling coefficients 
(which enforce $m_1+\ldots m_{N-1} = 0$) and $M\equiv \{m_1\ldots m_{N-1}\}$. The coupling coefficients can be written either in terms of Clebsch-Gordan coefficients or Wigner 3-$j$ symbols \citep[e.g.,][\S34.2]{nist_dlmf};
in the latter case, we obtain
\beq\label{eq: basis-coupling}
    C^\L_M \equiv C^{\ell_1\ell_2(\ell_{12})\ldots\ell_{N-1}}_{m_1m_2\ldots m_{N-1}} &=& \mathcal{E}(\L)\sqrt{2\ell_{12}+1}\times\cdots\times\sqrt{2\ell_{12..N-3}+1}\\\nonumber
    &&\,\times\,\sum_{m_{12}\ldots}(-1)^\kappa \tj{\ell_1}{\ell_2}{\ell_{12}}{m_1}{m_2}{-m_{12}}\tj{\ell_{12}}{\ell_3}{\ell_{123}}{m_{12}}{m_3}{-m_{123}}\ldots\tj{\ell_{12\ldots N-3}}{\ell_{N-2}}{\ell_{N-1}}{m_{12\ldots N-3}}{m_{N-2}}{m_{N-1}},
\eeq
with $\mathcal{E}(\L) = (-1)^{\sum_i \ell_i}$ and $\kappa = \ell_{12}-m_{12}+\ell_{123}-m_{123}+\ldots +\ell_{12\ldots N-3}-m_{12\ldots N-3}$. The basis \eqref{eq: basis-def} can be derived by considering the eigenstates of the angular momentum operators; in particular the basis is an eigenstate of $\vec{L}_1$ with eigenvalue $\ell_1(\ell_1+1)$, an eigenstate of $(\vec{L}_1+\vec{L}_2)^2$ with eigenvalue $\ell_{12}(\ell_{12}+1)$ \textit{et cetera}, as well as a zero-eigenvalue eigenstate of the combined angular momentum operator $(\vec L_1+\ldots \vec L_{N-1})^2$. We note that this basis is simply related to the polypolar spherical harmonics of \citet{1988qtam.book.....V}, except that we restrict to zero total angular momentum. Furthermore, the $N=4$ case is a rescaled version of a subset of the tripolar spherical harmonics, which have been used a number of times in cosmological contexts \citep[e.g.,][]{2017PhRvD..95f3508S}.

Comparison to \eqref{eq: isotropic-basis-expansion} shows that the basis functions are parametrized by sets of integers $\L = \{\ell_1,\ell_2,(\ell_{12}),\ell_3,(\ell_{123}),\ldots,\ell_{N-1}\}$, hereafter denoted \textit{multiplets}. These contain two types of contributions: (a) principal angular momenta $\ell_i$ (which index the spherical harmonics), and (b) intermediate angular momenta $\ell_{12}$, $\ell_{123}$ \textit{et cetera} (appearing only in the coupling $C_M^\L$).\footnote{Throughout this work, we indicate intermediate angular momentum components by parentheses.} For $N>4$, intermediates are required since there is more than one way to combine the $(N-1)$ spherical harmonics to obtain a state of zero combined angular momentum; put another way, due to the vectorial-nature of angular momentum addition, the principal angular momenta $\{\ell_1,\ell_2,\ldots,\ell_{N-1}\}$ are insufficient to fully define the state. From \eqref{eq: basis-coupling}, we note several constraints on the values of $\ell$ and $m$. In particular, the 3-$j$ symbols enforce that $m_{12}=m_1+m_2$, $m_{123} = m_{12}+m_3$ \textit{et cetera} and thus that $\sum_i m_i = 0$, giving invariance under $z$-axis rotations. This ensures that the basis function \eqref{eq: basis-def} is a state of zero combined orbital angular momentum. Furthermore, each 3-$j$ symbol implies a triangle inequality of the form $|\ell_1-\ell_2|\leq \ell_{12}\leq \ell_{1}+\ell_2$. This restricts the number of physical $\L$ states.

\subsection{Basic Properties}\label{subsec: basis-properties}
The 
coupling coefficients, $C^\L_M$, appearing in \eqref{eq: basis-def} satisfy: 
\beq\label{eq: basis-coupling-M-inversion}
    C^\L_{-M} = \mathcal{E}(\L)\;C^\L_M ,
\eeq
where $-M = \{-m_1,-m_2,\ldots -m_{N-1}\}$, and $\mathcal{E}(\L) = +1$ if the sum of all principal angular momentum ($\sum_i\ell_i$) is even, and $-1$ else. Consequently the isotropic basis functions obey the conjugate condition
\beq\label{eq: basis-conjugation}
    \P_\L^*(\hR) = \mathcal{E}(\L)\P_\L(\hR)
\eeq
\citep{2020arXiv201014418C}. Under parity transformations, $\mathbb{P}[Y_{\ell m}(\hr)] \equiv Y_{\ell m}(-\hr) = (-1)^\ell Y_{\ell m}(\hr)$, with parity operator $\mathbb{P}$. Thus
\beq\label{eq: basis-parity-relation}
    \mathbb{P}[\P_\L](\hR) = \mathcal{E}(\L)\P_\L(\hR).
\eeq
Due to this property, we will refer to the cases $\mathcal{E}(\L) = +1$ and $-1$ as \textit{parity-even} and \textit{parity-odd} respectively. We further note that \eqref{eq: basis-conjugation} implies that $\P_\L(\hR)$ is purely real for even-parity $\L$ and purely imaginary for odd-parity $\L$.

Secondly, the basis functions are orthonormal when integrated over $\hR$:
\beq\label{eq: basis-orthogonality}
    \int d\hR\,\P_\L(\hR)\P^*_{\L'}(\hR) = \delta^\mathrm{K}_{\L\L'}\,,
\eeq
where $d\hR = d\hr_1\ldots d\hr_{N-1}$ and the Kronecker delta is unity if all components of $\L$ and $\L'$ match and zero else. This follows straightforwardly from \eqref{eq: basis-coupling} and the orthonormality of the spherical harmonics. Given the expansion \eqref{eq: isotropic-basis-expansion}, this allows the basis coefficients $f_\L$ to be written as
\beq\label{eq: basis-coeff-extraction}
    f_\L(R) &=& \int d\hR\,f(\vR)\P^*_\L(\hR),
\eeq
adopting the notation $R = \{r_1,\ldots ,r_{N-1}\}$. For $\L = \vec 0$, the basis functions take the simple form
\beq\label{eq: basis-zero-ell}
    \P_{\vec 0}(\hR) = (4\pi)^{-(N-1)/2},
\eeq
using $Y_{00}(\hr) = (4\pi)^{-1/2}$ and $C^{\vec 0}_{\vec 0}=1$.

An additional property of interest is the contraction of two basis functions of the same arguments. This is the analog of the well-known spherical harmonic product-to-sum relation;
\beq
    Y_{\ell_1m_1}(\hr)Y_{\ell_2m_2}(\hr) = \sum_{\ell_3m_3}\mathcal{G}^{\ell_1\ell_2\ell_3}_{m_1m_2(-m_3)}(-1)^{m_3}Y_{\ell_3m_3}(\hr)
\eeq
\citep[Eq.\,34.3.22]{nist_dlmf}, where $\mathcal{G}^{\ell_1\ell_2\ell_3}_{m_1m_2(-m_3)}$ is the Gaunt integral, defined by
\beq
    \mathcal{G}^{\ell_1\ell_2\ell_3}_{m_1m_2m_3} &\equiv& \left[\frac{(2\ell_1+1)(2\ell_2+1)(2\ell_3+1)}{4\pi}\right]^{1/2}\tj{\ell_1}{\ell_2}{\ell_3}{m_1}{m_2}{m_3}\tjo{\ell_1}{\ell_2}{\ell_3}.
\eeq
In our context, we find
\beq\label{eq: basis-function-product}
    \P_\L(\hR)\P_{\L'}(\hR)&=& \sum_{\L''}\mathcal{E}(\L'')\mathcal{G}^{\L\L'\L''}\P_{\L''}(\hR)
\eeq
\citep[\S6]{2020arXiv201014418C}, where $\mathcal{G}^{\L\L'\L''}$ is the generalized Gaunt integral, defined as
\beq\label{eq: generalized-gaunt-def}
    \mathcal{G}^{\L\L'\L''} = \int d\hR\,\P_{\L}(\hR)\P_{\L'}(\hR)\P_{\L''}(\hR).
\eeq
In Appendix \ref{appen: generalized-gaunt}, we provide expressions for $\mathcal{G}^{\L\L'\L''}$ with $N=3,4,$ and $5$, involving Wigner 9-$j$ symbols.

\subsection{Special Cases}\label{subsec: basis-def}
For later use, we give the explicit forms of the basis functions for $N=3,4,5$. For $N=3$, the basis is specified by two components $\L = \{\ell_1,\ell_2\}$, with 
\beq\label{eq: N=3-coeff}
    C^\L_M = C^{\ell_1\ell_2}_{m_1m_2} = \frac{(-1)^{\ell_1-m_1}}{\sqrt{2\ell_1+1}}\delta^\mathrm{K}_{\ell_1\ell_2}\delta^\mathrm{K}_{m_1-m_2},
\eeq
which enforces $\ell_1=\ell_2$, $m_1+m_2=0$.\footnote{Note that these are equivalent to the $N=2$ basis functions of \citet{2020arXiv201014418C}.} From \eqref{eq: basis-def}, the full basis function can thus be written
\beq\label{eq: N=3-basis}
    \P_{\ell}(\hr_1,\hr_2) = \sum_{m=-\ell}^\ell\frac{(-1)^{\ell-m}}{\sqrt{2\ell+1}}Y_{\ell m}(\hr_1)Y_{\ell-m}(\hr_2) &=& \frac{\sqrt{2\ell+1}}{4\pi}(-1)^\ell L_\ell(\hr_1\cdot\hr_2),
\eeq
where $L_\ell(\mu)$ is the Legendre polynomial of order $\ell$. In this case, our basis is simply a rescaled version of a Legendre expansion into multipoles. Note that $\ell_1+\ell_2 \equiv 2\ell$ is always even, thus all $N=3$ basis functions are parity-even. This occurs since the $N=3$ parity transformation is equivalent to a rotation, and the basis functions are invariant under rotations.

For $N=4$, simplifying the coefficients of \eqref{eq: basis-coupling} gives
\beq\label{eq: N=4-coeff}
    C^\L_M = C^{\ell_1\ell_2\ell_3}_{m_1m_2m_3} = (-1)^{\ell_1+\ell_2+\ell_3}\tj{\ell_1}{\ell_2}{\ell_3}{m_1}{m_2}{m_3},
\eeq
depending on three principal angular momenta, $\L = \{\ell_1,\ell_2,\ell_3\}$. The full basis function becomes
\beq\label{eq: N=4-basis}
    \P_{\ell_1\ell_2\ell_3}(\hr_1,\hr_2,\hr_3) = (-1)^{\ell_1+\ell_2+\ell_3}\sum_{m_1m_2m_3}\tj{\ell_1}{\ell_2}{\ell_3}{m_1}{m_2}{m_3}Y_{\ell_1m_1}(\hr_1)Y_{\ell_2m_2}(\hr_2)Y_{\ell_3m_3}(\hr_3).
\eeq
As before this is just a simple combination of spherical harmonics with the coupling chosen to ensure rotational invariance. Here, the sum of the principal angular momenta, $\ell_1+\ell_2+\ell_3$, can be either even or odd, thus both even- and odd-parity functions are permitted.

Finally, we remark on the form for $N=5$. Here, we require \textit{five} sets of angular momentum coefficients to fully specify each basis function, $\L = \{\ell_1,\ell_2,(\ell_{12}),\ell_3,\ell_4\}$. An intermediate angular momentum $\ell_{12}$ appears as there are multiple ways to combine four spherical harmonics with given principal angular momenta $\ell_i$ such that the total angular momentum is zero. In this case, the coupling coefficients and associated basis functions are
\beq\label{eq: N=5-basis-and-coeff}
    C^\L_M = C^{\ell_1\ell_2(\ell_{12})\ell_3\ell_4}_{m_1m_2m_3m_4} &=& (-1)^{\ell_1+\ell_2+\ell_3+\ell_4}\sum_{m_{12}}(-1)^{\ell_{12}-m_{12}}\tj{\ell_1}{\ell_2}{\ell_{12}}{m_1}{m_2}{-m_{12}}\tj{\ell_{12}}{\ell_3}{\ell_4}{m_{12}}{m_3}{m_4},\\\nonumber
    \P_{\ell_1\ell_2(\ell_{12})\ell_3\ell_4}(\hr_1,\hr_2,\hr_3,\hr_4) &=& (-1)^{\ell_1+\ell_2+\ell_3+\ell_4}\sum_{m_{12}}(-1)^{\ell_{12}-m_{12}}\sum_{m_1m_2m_3m_4}\tj{\ell_1}{\ell_2}{\ell_{12}}{m_1}{m_2}{-m_{12}}\tj{\ell_{12}}{\ell_3}{\ell_4}{m_{12}}{m_3}{m_4}\\\nonumber
    &&\,\times\,Y_{\ell_1m_1}(\hr_1)Y_{\ell_2m_2}(\hr_2)Y_{\ell_3m_3}(\hr_3)Y_{\ell_4m_4}(\hr_4).
\eeq
Higher orders proceed similarly, with the basis function of $(N-1)$ coordinates involving $(N-4)$ intermediate angular momenta.

\section{The NPCF Estimator}\label{sec: estimator}

\subsection{Definition}\label{subsec: estimator-def}

The \resub{3D} $N$-point correlation function is defined as a statistical average over the product of $N$ overdensity fields, $\delta$ \resub{in some redshift bin}: 
\beq\label{eq: NPCF-def}
    \zeta(\vr_1,\ldots ,\vr_{N-1}) &=& \av{\delta(\vs)\delta(\vs+\vr_1)\ldots \delta(\vs+\vr_{N-1})}\\\nonumber
    &=&\int \frac{d\vs}{V}\;\delta(\vs)\delta(\vs+\vr_1)\ldots \delta(\vs+\vr_{N-1}),
\eeq
using the ergodic theorem to convert the statistical average into a spatial integral in the second line, assuming the volume $V$ to be large. By statistical homogeneity, the NPCF cannot depend on the absolute position $\vs$, thus it is a function of only $(N-1)$ sets of coordinates. Further, if one assumes $\delta$ to be statistically isotropic, the function must be invariant under any global rotation, \textit{i.e.} one in which all separation vectors $\hr$ are simultaneously rotated about some common origin. This is exactly the condition for the function to be expressed in the isotropic basis of \S\ref{sec: basis}; thus we may write
\beq\label{eq: NPCF-basis-decomposition}
    \boxed{\zeta(\vr_1,\ldots ,\vr_{N-1}) = \sum_\L \zeta_\L(r_1,\ldots ,r_{N-1})\P_\L(\hr_1,\ldots ,\hr_{N-1})}
\eeq
without loss of generality, where $\P_\L$ are the isotropic basis functions of $(N-1)$ coordinates defined in \S\ref{eq: basis-def}. $\L$ specifies the basis function, and is composed of angular momentum coefficients $\ell_1,\ell_2,\ell_{12},\ldots$, which take integer values. \resub{The coefficients indicate the scales of the \textit{internal} angles of the $N$-hedron of particles forming the NPCF (with $\ell_i$ corresponding to the side with length $r_i$).} For $N=3$, inserting the explicit basis function of \eqref{eq: N=3-basis} gives
\beq
    \zeta(\vr_1,\vr_2) &=& \sum_{\ell}\frac{\sqrt{2\ell+1}}{4\pi}(-1)^\ell \zeta_\ell(r_1,r_2)L_\ell(\hr_1\cdot\hr_2),
\eeq
thus the 3PCF basis coefficients $\zeta_\ell$ are simply rescaled Legendre multipoles (matching \citealt{2015MNRAS.454.4142S}). 

\resub{The above equations involve a sum over an infinite number of $\L$ coefficients. However, provided that we analyze the basis coefficients directly (rather than using them to reconstruct the full $\zeta$), it is fully valid to restrict to a subset of $\Lambda$ values with $\ell_i\leq \ell_{\rm max}$, given that the basis functions are orthogonal \eqref{eq: basis-orthogonality}. Here, $\ell_{\rm max}$ sets the resolution of the internal $N$-hedron angles; for example, restricting to $\ell_{\rm max} = 10$ allows us to resolve opening angles of $\approx 180^\circ/10 = 18^\circ$. Since all $N$ points in the polyhedron are located at similar distances from the observer, there is no need for $\ell_{\rm max}$ to be particularly large; this differs from analysis of the CMB and photometric galaxy samples, where one point of the $N$-hedron is located at the observer. In the latter case, $(N-1)$ side lengths are large (equal to the distance to the sample), thus the $N$-hedra are squeezed and a high $\ell_{\rm max}$ is typically required.}

As noted in \S\ref{subsec: basis-properties}, the basis functions come in two forms; parity-odd (with even $\ell_1+\ldots+\ell_{N-1}$) and parity-even (with odd $\ell_1+\cdots+\ell_{N-1}$). To understand our notation, first consider the NPCF under parity transforms:
\beq
    \mathbb{P}\left[\zeta(\vr_1,\ldots,\vr_{N-1})\right] &=& \zeta(-\vr_1,\ldots,-\vr_{N-1})\\\nonumber
    &=& \sum_\L \zeta_\L(r_1,\ldots,r_{N-1})\P_\L(-\hr_1,\ldots,-\hr_{N-1}) \equiv \sum_\L \mathcal{E}(\L)\zeta_\L(r_1,\ldots,r_{N-1})\P_\L(\hr_1,\ldots,\hr_{N-1}),
\eeq
where $\mathbb{P}$ is the parity operator and we have used the basis decomposition \eqref{eq: NPCF-basis-decomposition} and the parity inversion formula \eqref{eq: basis-parity-relation} in the second line. Invoking orthogonality \eqref{eq: basis-orthogonality}, we find that multiplets $\zeta_\L$ with odd $\ell_1+\ldots+\ell_{N-1}$ are odd under parity-inversion, whilst those with even $\ell_1+\ldots+\ell_{N-1}$ are even under parity-inversion. Furthermore, given that the odd-parity basis functions are imaginary (\S\ref{subsec: basis-properties}), this implies that all $\zeta_\L$ with odd-parity $\L$ are also purely imaginary (as found in Appendix \ref{appen: spin-sum}). Assuming the NPCF to be parity-symmetric, we expect $\zeta_\L$ to vanish for odd $\ell_1+\ldots+\ell_{N-1}$. We further note that odd-parity terms appear only for $N>3$, as noted in \S\ref{subsec: basis-properties}. This occurs since parity inversion is equivalent to a 3D rotation for $N\leq 3$, and the isotropic NPCF is rotationally invariant.

Finally, we note that the general NPCF is a sum of two contributions: `connected' and `disconnected'. The first involves the joint probability of $N$ points in space, whilst the second factorizes into multiple independent components. For example, the cosmological 4PCF contains contributions both from the intrinsic non-Gaussian 4PCF (the connected piece), and the product of two 2PCFs (the disconnected piece). In full:
\beq\label{eq: 4pcf-discon-con}
    \zeta(\vr_1,\vr_2,\vr_3) &=& \av{\delta(\vs)\delta(\vs+\vr_1)\delta(\vs+\vr_2)\delta(\vs+\vr_3)}_c + \left[\av{\delta(\vs)\delta(\vs+\vr_1)}\av{\delta(\vs+\vr_2)\delta(\vs+\vr_3)} + \text{2 perms.}\right]\\\nonumber
    &=&\zeta_c(\vr_1,\vr_2,\vr_3)+\left[\xi(\vr_1)\xi(\vr_2-\vr_3)+\text{ 2 perms.}\right],
\eeq
where the subscript `c' indicates a connected contribution and $\xi(\vr)$ is the 2PCF. Similarly, the 6PCF contains a connected non-Gaussian term and disconnected terms involving (a) two 3PCFs, (b) a 4PCF and a 2PCF, and (c) three 2PCFs. A detection of the NPCF signal is thus not necessarily a detection of non-Gaussianity; most of the signal-to-noise is expected to be in the disconnected contribution, which is degenerate with the 2PCF for even $N$. To measure the \textit{connected} part we must subtract off the disconnected contribution either at the estimator-level or using some theory model; this will be discussed in depth in \citet{4pcf_boss} for the 4PCF.

\subsection{General Estimator}\label{subsec: estimator-continuous}

To extract the NPCF multiplets $\zeta_\L(r_1,\ldots ,r_{N-1})$, we may use the orthogonality relation \eqref{eq: basis-coeff-extraction}:
\beq\label{eq: extract-npcf-mult}
    \zeta_\L(r_1,\ldots ,r_{N-1}) &=& \int d\hr_1\ldots d\hr_{N-1}\,\zeta(\vr_1,\vr_2,\ldots ,\vr_{N-1})\P^*_\L(\hr_1,\hr_2,\ldots ,\hr_{N-1})\\\nonumber
    &=& \mathcal{E}(\L)\int d\hr_1\ldots d\hr_{N-1}\,\zeta(\vr_1,\vr_2,\ldots ,\vr_{N-1})\P_\L(\hr_1,\hr_2,\ldots ,\hr_{N-1}),
\eeq
using the conjugation rule \eqref{eq: basis-conjugation} to obtain the second line. Inserting the NPCF definition \eqref{eq: NPCF-def} and the explicit forms of the basis functions \eqref{eq: basis-def} gives an estimator for $\zeta_\L$:
\beq\label{eq: NPCF-mult-estimator-tmp}
    \hat\zeta_\L(r_1,\ldots ,r_{N-1}) = \frac{1}{V}\int d\vs \sum_{m_1\ldots m_{N-1}}\mathcal{E}(\L)\;C^\L_{M}\int d\hr_1\ldots d\hr_{N-1}\;\delta(\vs)\delta(\vs+\vr_1)\ldots \delta(\vs+\vr_{N-1})Y_{\ell_1m_1}(\hr_1)\ldots Y_{\ell_{N-1}m_{N-1}}(\hr_{N-1}),
\eeq
where we have integrated the absolute position $\vs$ over a volume $V$, as in \eqref{eq: NPCF-def}. Note that we have not yet introduced radial binning. The appeal of this decomposition is clear; the estimator is \textit{exactly separable in $\hr_i$}. Explicitly, we may define
\beq\label{eq: alm-def}
    a_{\ell m}(\vs;r) = \int d\hr\,\delta(\vs+\vr)Y_{\ell m}(\hr)
\eeq
(which is simply a convolution of $\delta$ and $Y_{\ell m}$), and write \eqref{eq: NPCF-mult-estimator-tmp} as
\beq\label{eq: NPCF-mult-estimator}
    \boxed{\hat\zeta_\L(r_1,\ldots ,r_{N-1}) = \frac{1}{V}\int d\vs\,\delta(\vs)\sum_{m_1\ldots m_{N-1}}\mathcal{E}(\L)\;C^\L_{M}a_{\ell_1m_1}(\vs;r_1)\ldots a_{\ell_{N-1}m_{N-1}}(\vs;r_{N-1}).}
\eeq
Given the harmonic coefficients $a_{\ell m}(\vs;r)$, the NPCF is formed by summing over the $m$ indices, weighted by the coupling coefficients $\mathcal{E}(\L)\;C^\L_{M}$, and integrating over position $\vs$. Restricting to $\ell_i\leq \ell_\mathrm{max}$ \resub{(and thus bounding the resolution of the internal $N$-hedron angles)}, an upper bound to the number of summation terms is given by $(1+2\ell_\mathrm{max})^{N-2}$. Whilst \eqref{eq: NPCF-mult-estimator} is unbiased (as shown by inserting \ref{eq: NPCF-def} and \ref{eq: NPCF-basis-decomposition}, then using orthonormality), it is not strictly optimal, \textit{i.e.} it is not minimum variance. Optimal estimation of the higher-point functions is possible (and discussed for example in \citet{2011MNRAS.417....2S,2015arXiv150200635S,2011arXiv1105.2791F,2012PhRvD..86f3511F,bk_opt} for Fourier-space statistics); this generally involves extra additive terms. We do not explore such complications in this work.

In practice, one usually bins the NPCFs in radius. To do so, we introduce a binning function $\Theta^b(r)$ which is unity if $r$ is in \resub{radial} bin $b$ and zero else. In this case, the \resub{radially-averaged} NPCF becomes
\beq
    \zeta^B_\L &=& \frac{1}{v_B}\int \left[\prod_{i=1}^{N-1}r_i^2dr_i\,\Theta^{b_i}(r_i)\right]\zeta_\L(r_1,\ldots ,r_{N-1}),\\\nonumber
    v_B &=& \int \left[\prod_{i=1}^{N-1}r_i^2dr_i\,\Theta^{b_i}(r_i)\right],
\eeq
where the second line gives the bin volume. We denote the set of all \resub{radial} bin indices by $B = \{b_1,...,b_{N-1}\}$, with $b_i$ referring to the $N$-hedron side with radius $r_i$ and thus angular momentum coefficient $\ell_i$. 
Since the binning is separable in $r_i$, the corresponding NPCF estimator may be written
\beq\label{eq: NPCF-mult-binned-estimator}
    \boxed{\hat\zeta^B_\L = \frac{1}{V}\int d\vs\,\delta(\vs)\sum_{m_1\ldots m_{N-1}}\mathcal{E}(\L)\;C^\L_{M}a^{b_1}_{\ell_1m_1}(\vs)\cdots a_{\ell_{N-1}m_{N-1}}^{b_{N-1}}(\vs),}
\eeq
with the new harmonic coefficients
\beq\label{eq: alm-binned-def}
    a^{b}_{\ell m}(\vs) = \frac{1}{v_b}\int d\vr\,\delta(\vs+\vr)Y_{\ell m}(\hr)\Theta^b(r)\,,
\eeq
where $v_b \equiv \int d\vr\,\Theta^b(r)$ is the bin volume. This is again a simple convolution of $\delta$ and $Y_{\ell m}\Theta^b$.
In this approach, binning has an important function; it allows us to estimate only a finite number of $a_{\ell m}^{b}(\vs)$ harmonic coefficients, before performing the summation of \eqref{eq: NPCF-mult-binned-estimator}.

\subsubsection{Discrete Data}\label{subsec: estimator-discrete}
In most contexts, we are interested in the estimation of NPCFs from discrete data. Given a set of $\Ng$ particle positions at $\{\vr_i\}$, the discrete density field may be written as a sum of Dirac delta functions and weights $w_i$:
\beq
    \delta(\vr) = \sum_{i=1}^{\Ng} w_i\;\delta_\mathrm{D}(\vr-\vr_i).
\eeq
Using this, the $\vs$ integral in the NPCF estimator becomes a sum over particle positions as
\beq\label{eq: NPCF-mult-discrete-binned-estimator}
    \boxed{\hat\zeta^B_\L = \frac{1}{V}\sum_i w_i \sum_{m_1\ldots m_{N-1}}\mathcal{E}(\L)\;C^\L_{M}a^{b_1}_{\ell_1m_1}(\vs_i)\cdots a_{\ell_{N-1}m_{N-1}}^{b_{N-1}}(\vs_i),}
\eeq
and the 
harmonic coefficients $a_{\ell m}^b$ can be written
\beq\label{eq: binned-discrete-alm}
    \boxed{a_{\ell m}^{b}(\vs_i) = \frac{1}{v_b}\sum_j w_j Y_{\ell m}(\widehat{\vec r_j-\vs_i}) \Theta^b(|\vr_j-\vs_i|).}
\eeq
Note that the $a_{\ell m}^b$ harmonic coefficients 
need to be evaluated only at the locations of the particles, $\{\vr_i\}$. The interpretation of \eqref{eq: binned-discrete-alm} is straightforward; for a given primary position $\vs_i$, we sum over all secondary positions $\vr_j$ which lie within bin $b$ of $\vs_i$, weighted by the spherical harmonics evaluated at the separation vector $\vr_j-\vs_i$. Practically this is evaluated as a weighted sum of pairs of particles, \textit{i.e.} a \textit{pair-count}. 

The full procedure for estimating the binned NPCF from a discrete set of particles is thus:
\begin{enumerate}
    \item For a given primary particle $i$, compute $a_{\ell m}^b(\vs_i)$ as a weighted sum of spherical harmonics.
    \item For each $\L$ multiplet and bin combination $B$, sum the product of $(N-1)$ harmonic coefficients over all $m_i$, weighting by the coupling coefficient $\mathcal{E}(\L)\;C^\L_{M}$.
    \item Repeat for each primary particle and sum.
\end{enumerate}
Since it is a sum over secondary positions $\vr_j$, computation of \eqref{eq: binned-discrete-alm} scales as $\Ng$ for total particle number $\Ng$. Given that each $a_{\ell m}^b$ function must be evaluated at the location of each particle, the algorithm has complexity $\mathcal{O}(\Ng^2)$. This is the same as for the 3PCF \citep{2015MNRAS.454.4142S}, and \textit{much} better than the na\"ive $\mathcal{O}(\Ng^{N})$ complexity expected from counting every possible $N$-tuplet of particles and assigning them to bins. \resub{Additional details of the computational scalings can be found in \S\ref{subsec: scalings}.}

For discrete data, an important subtlety should be noted. If the bins $B = \{b_1,\ldots ,b_{N-1}\}$ are not disjoint (\textit{i.e.} we include scenarios in which multiple side-lengths of the $N$-hedron fall into the same radial bin), the same secondary particle can contribute to multiple $a_{\ell m}^{b}$ harmonic coefficients. Physically this corresponds to a scenario in which we have an $N$-point function with multiple coincident points. As discussed in \citet{2021MNRAS.501.4004P}, these `self-count' contributions can be removed by subtracting a hierarchy of lower $N$-point functions.\footnote{Specifically, we must subtract off basis functions with two coincident points, e.g., $\P_\Lambda(\vr_1,\vr_1,\ldots ,\vr_{N-2})$, those with three coincident points, e.g., $\P_\Lambda(\vr_1,\vr_1,\vr_1,\ldots ,\vr_{N-3})$, \textit{et cetera}. Using the contraction formulae of \citet[\S5]{2020arXiv201014418C}, these can be written as lower-order basis functions.} However, given that these NPCF regimes are difficult to model and interpret, a simpler approach is to enforce that all $b_i$ \resub{radial} bins are distinct, effectively enforcing some minimum separation on the side lengths of the $N$-hedron. In this work, we assume $b_1<b_2<\ldots <b_{N-1}$ for this reason.

\subsubsection{Gridded Data}
Estimator \eqref{eq: NPCF-mult-binned-estimator} may be similarly applied to data discretized onto a grid, for example a matter density field in an $N$-body simulation, or some fluid in a hydrodynamic simulation. 
If the density field $\delta$ is defined on some grid, we may compute the harmonic coefficients using a Fast Fourier Transform (FFT):
\beq\label{eq: alm-binned-def-fft}
    a^{b}_{\ell m}(\vs) = \ift{\ft{\delta}(\vk)\ft{Y_{\ell m}(\hr)\Theta^b(r)}(\vk)}(\vs)\left/ \left[\int d\vr\,\Theta^b(r)\right]\right.,
\eeq
where $\mathcal{F}$ ($\mathcal{F}^{-1}$) is the forward (reverse) Fourier operator. Assuming $N_\mathrm{FFT}$ grid-points, this has complexity $\mathcal{O}(N_\mathrm{FFT}\log N_\mathrm{FFT})$ and allows estimation of $a_{\ell m}^b$ in all grid-cells simultaneously.\footnote{We caution that this approach requires many $a_{\ell m}^b(\vs)$ fields must be stored in order to compute \eqref{eq: NPCF-mult-binned-estimator}; this results in significant memory usage.} Practically, we may use \eqref{eq: NPCF-mult-binned-estimator} to compute the NPCF of a gridded field $\delta$ in the following fashion:
\begin{enumerate}
    \item For each $\ell, m$ up to some $\ell_\mathrm{max}$ and radial bin $b$, estimate $a_{\ell m}^b(\vs)$ using FFTs.
    \item Take the product of $N-1$ $a_{\ell m}^b$ fields and integrate over space, weighted by $\delta(\vs)$.
    \item Sum over $M = \{m_1,\ldots, m_{N-1}\}$, weighted by the coupling coefficients $\mathcal{E}(\L)\;C^\L_{M}$.
\end{enumerate}
This generalizes the approach of \citet{2018ApJ...862..119P}, which computed the 3PCF of gridded data via Fourier methods. Whilst this approach may prove useful for the analysis of simulations and applications beyond cosmology, we focus on discrete data for the remainder of this work, though much of the discussion applies to both cases.

\subsection{Edge Correction}\label{subsec: edge-correction}
In the analysis of real galaxy surveys, one does not have direct access to the galaxy overdensity field $\delta$. Instead, our observables are the set of $N_D$ galaxy positions $\{\vr_i\}$ (hereafter `data'), as well as a set of $N_R$ unclustered particles (hereafter `randoms') $\{\vr_j\}$, with the randoms distributed according to the survey geometry and selection function. Mathematically, the associated random fields are given by
\beq\label{eq: data-randoms-def}
    D(\vr) &=& \sum_{i=1}^{N_D} w^D_i\delta_\mathrm{D}(\vr-\vr_i), \quad R(\vr) = \sum_{j=1}^{N_R} w^R_j\delta_\mathrm{D}(\vr-\vr_j),
\eeq
where $w^D_i$ and $w^R_i$ are data and random weights, with the latter normalized such that $\sum_i w_i^D = \sum_j w_j^R$. Averaging over the (Poissonian) sampling distribution, the data and random fields have expectations $\av{D(\vr)} = n(\vr)[1+\delta(\vr)]$, $\av{R(\vr)} = n(\vr)$ where $n(\vr)$ is some background number density. This motivates the following estimator for the galaxy NPCF, based on \citet{1993ApJ...412...64L} and \citet{1998ApJ...494L..41S}:
\beq\label{eq: generalized-landy-szalay}
    \hat\zeta(\vr_1,\ldots ,\vr_{N-1}) = \frac{\int d\vs\,[D-R](\vs)[D-R](\vs+\vr_1)\ldots [D-R](\vs+\vr_{N-1})}{\int d\vs\,R(\vs)R(\vs+\vr_1)\ldots R(\vs+\vr_{N-1})} \equiv  \frac{\mathcal{N}(\vr_1,\ldots ,\vr_{N-1})}{\mathcal{R}(\vr_1,\ldots ,\vr_{N-1})},
\eeq
defining the functions $\mathcal{N}$ and $\mathcal{R}$. In the shot-noise limit, this is simply an inverse-variance weighting \citep[cf.\,][]{2015MNRAS.454.4142S}. For a simple periodic-box geometry in the limit of infinite randoms, $R(\vr)$ takes the constant value $\bar{n}$ everywhere, thus the denominator is trivial. Note that \eqref{eq: generalized-landy-szalay} applies also to gridded density fields with non-uniform geometry; in this case $\mathcal{N}_\L$ and $\mathcal{R}_\L$ are the multiplets of the \textit{gridded} `data-minus-random' and random density fields.

In general, we can expand each function appearing in \eqref{eq: generalized-landy-szalay} in the isotropic basis of \S\ref{sec: basis}, giving
\beq
    \sum_\L\hat\zeta_\L(R)\P_\L(\hR) &=& \frac{\sum_{\L''}\mathcal{N}_{\L''}(R) \P_{\L''}(\hR)}{\sum_{\L'}\mathcal{R}_{\L'}(R)\P_{\L'}(\hR)},
\eeq
where we have denoted $\hR \equiv \{\hr_1,\ldots ,\hr_{N-1}\}$ and $R\equiv \{r_1,\ldots,r_{N-1}\}$. 
Defining $f_\L(R) \equiv \mathcal{R}_\L(R)/\mathcal{R}_{\vec 0}(R)$, we can write
\beq\label{eq: edge-correction-tmp}
    \sum_{\L''}\frac{\mathcal{N}_{\L''}(R)}{\mathcal{R}_{\vec 0}(R)}\P_{\L''}(\hR) &=& \sum_{\L\L'}\hat\zeta_\L(R) f_{\L'}(R)\P_{\L}(\hR)\P_{\L'}(\hR)\\\nonumber
    &=& \sum_{\L\L'}\hat\zeta_\L(R) f_{\L'}(R)\left[\sum_{\L''}\mathcal{E}(\L'')\mathcal{G}^{\L\L'\L''}\P_{\L''}(\hR)\right],
\eeq
where we have expanded the product of two isotropic basis functions using \eqref{eq: basis-function-product}. This uses the generalized Gaunt integrals $\mathcal{G}^{\L\L'\L''}$ defined in \eqref{eq: generalized-gaunt-def}, with explicit low-order forms given in terms of Wigner 3-$j$ and 9-$j$ symbols in Appendix \ref{appen: generalized-gaunt}. Invoking orthonormality \eqref{eq: basis-orthogonality}, we can extract the $\P_{\L''}(\hR)$ basis coefficient as
\beq\label{eq: edge-correction-matrix}
    \frac{\mathcal{N}_{\L''}(R)}{\mathcal{R}_{\vec 0}(R)} = \sum_{\L}\hat\zeta_\L\left[\sum_{\L'}f_{\L'}(R)\;\mathcal{E}(\L'')\mathcal{G}^{\L\L'\L''}\right] \equiv \sum_\L \hat\zeta_\L(R) M_{\L\L''}(R).
\eeq
In the second line we have noted that this is simply a linear equation with (radial-bin dependent) coupling matrix $M_{\L\L''}$; inverting yields the \textit{edge-corrected} estimator for $\hat\zeta_\L$:
\beq\label{eq: edge-corrected-NPCF}
    \boxed{\hat\zeta_\L(r_1,\ldots ,r_{N-1}) = \sum_{\L''}M^{-1}_{\L\L''}(r_1,\ldots ,r_{N-1})\frac{\mathcal{N}_{\L''}(r_1,\ldots ,r_{N-1})}{\mathcal{R}_{\vec 0}(r_1,\ldots ,r_{N-1})},}
\eeq
writing out $R$ in full. This applies also to the binned statistic with the substitutions $\mathcal{N}_\L(R)\to \mathcal{N}_\L^B$, $\mathcal{R}_\L(R)\to\mathcal{R}_\L^B$, $\zeta_\L(R)\to\zeta_\L^B$, as can be seen from first rewriting \eqref{eq: generalized-landy-szalay} in binned form.

The meaning of \eqref{eq: edge-corrected-NPCF} can be understood as follows. For a uniform periodic geometry (e.g., an $N$-body simulation or some gridded periodic data), $\mathcal{R}(\vr_1,\ldots ,\vr_{N-1})$ is independent of positions $\vr_i$, thus its only non-zero basis coefficient is
\beq\label{eq: R-periodic-form}
    \mathcal{R}_{\vec 0}(r_1,\ldots ,r_{N-1}) = \frac{\bar{n}^N}{V}\int d\vs\,\int d\hr_1\cdots d\hr_{N-1}\,\P_{\vec 0}(\hr_1,\ldots ,\hr_{N-1}) = (4\pi)^{-\resub{(N-1)/2}}\bar{n}^N,
\eeq
using \eqref{eq: basis-zero-ell}. In this instance, $f_{\L'} \propto \delta^\mathrm{K}_{\L'\vec 0}$, and $M_{\L\L''}\propto \delta^\mathrm{K}_{\L\L''}$, hence the matrix multiplication is trivial. In general, a non-uniform survey geometry induces a coupling between multiplets, implying that the na\"ive estimator $\mathcal{N}_\L/\mathcal{R}_{\vec 0}$ is insufficient.\footnote{The na\"ive estimator is analogous to the window-convolved galaxy power spectrum estimator of \citet{1994ApJ...426...23F}, or the CMB \textit{pseudo}-$C_\ell$ estimator.} In this case, the matrix $M_{\L\L''}$ corrects for the effect of the survey window function, ensuring that $\hat\zeta_\L$ is unbiased.

The edge-correction equation \eqref{eq: edge-corrected-NPCF} is straightforward to implement; one first measures the multiplets $\mathcal{N}_\L$ and $\mathcal{R}_\L$ using the estimators of \eqref{eq: NPCF-mult-estimator} or \eqref{eq: NPCF-mult-discrete-binned-estimator}, with $\delta$ replaced by $[D-R](\vr)$ or $R(\vr)$. Given the set of all measured multiplets, the matrix inversion is a straightforward operation, and may be done for each radial bin combination separately. In practice, $M_{\L\L''}(R)$ is close to diagonal (cf.\,Fig.\,\ref{fig: coupling-matrices}), thus truncation of the infinite matrix $M_{\L\L''}(R)$ at finite $\ell_\mathrm{max}$ is expected to yield an accurate estimate of $\zeta_\L(R)$ for all multiplets containing $\ell$ near $\ell_\mathrm{max}$. \resub{In practice, this is ensured by measuring $\mathcal{N}_\L$ and $\mathcal{R}_\L$ using an $\ell_{\rm max}$ that is one larger than that used for the output NPCF.}\footnote{An additional caveat concerns \textit{anisotropy} of the window function. Strictly speaking, couplings between the anisotropic window function and anisotropic galaxy distribution can generate an isotropic NPCF, and ought to be removed via an anisotropic edge-correction equation. This is analogous to the mixing of power spectrum monopole and quadrupole in the presence of an anisotropic survey window. Here, we assume this effect to be small, such that the bias in our estimates of the isotropic functions can be ignored.} We additionally note that $\mathcal{G}^{\L\L'\L''}$ is only non-zero if $\sum_i \ell^{}_i+\sum_j \ell'_j+\sum_k\ell''_k$ is even. This implies that a parity-odd NPCF can be sourced by a parity-even $\mathcal{R}$ and a parity-odd $\mathcal{N}$ or \textit{vice versa}, whilst a parity-even NPCF requires both the data and random counts to have the same parity (cf.\,\S\ref{subsec: estimator-def}).

\section{Implementation}\label{sec: implementation}
Below, we discuss the implementation of the $\mathcal{O}(\Ng^2)$ discrete NPCF estimator given in \eqref{eq: NPCF-mult-discrete-binned-estimator}. Discussion of the gridded NPCF estimator (which is of use for hydrodynamic simulations and gridded cosmological simulations) will follow in future work, building upon the 3PCF code of \citet{2018ApJ...862..119P}. The discrete estimator is implemented within the code package, \textsc{encore}, publicly available at \href{https://github.com/oliverphilcox/encore}{github.com/oliverphilcox/encore}, and is loosely based on the isotropic 3PCF code of \citet{2015MNRAS.454.4142S} and the \textsc{hipster} power spectrum estimator package of \citet{2020MNRAS.492.1214P,2021MNRAS.501.4004P}. \textsc{encore} is written in C\textsc{++} and \textsc{cuda} and includes routines for computing the NPCF counts using both parallelized CPU resources and GPUs, as well as computing the edge-corrections as in \S\ref{subsec: edge-correction}. Currently, the code supports computation of the isotropic 2-, 3-, 4-, 5- and 6-point functions for both odd- and even-parity multiplets. Higher $N$ and $\ell_\mathrm{max}$ can be straightforwardly included, though we note that the $C^\Lambda_M$ matrix becomes sizeable for large $\ell_\mathrm{max}$ (scaling as $(1+\ell_\mathrm{max})^{N-1}(1+2\ell_\mathrm{max})^{2N-6}$, cf.\,\resub{\S\ref{subsec: scalings}}). 
The code can be run in various modes, as shown in Table \ref{tab: encore-modes} and discussed in greater detail below.

\begin{table}
    \centering
    \begin{tabular}{l|l}
    \textsc{fourpcf} & Compute the 4PCF\\
    \textsc{fivepcf} & Compute the 5PCF\\
    \textsc{sixpcf} & Compute the 6PCF\\
    \textsc{allparity} & Include both odd- and even-multiplets, rather than just even\\
    \textsc{disconnected} & Compute the disconnected (Gaussian) 4PCF\\
    \textsc{periodic} & Implement periodic boundary conditions (applicable for $N$-body simulations)\\
    \textsc{avx} & Accelerate the $a_{\ell m}^b$ computation \eqref{eq: alm-binned-def} using Advanced Vector Extensions (AVX) instruction sets\\
    \textsc{openmp} & Parallelize the main computation (steps 3-12 in Algorithm \ref{algo: overview}) using \textsc{OpenMP} on as many threads as are available\\
    \textsc{gpu} & Perform the $a_{\ell m}^{b}$ summation (Algorithm \ref{algo: summations}) on a Graphics Processing Unit (GPU) rather than a CPU\\
    \end{tabular}
    \caption{The various computation modes available in the \textsc{encore} NPCF package. These allow control over which statistics are computed, as well as architecture choices. Any combination of these can be used in tandem, except for \textsc{openmp} and \textsc{gpu}. In the \textsc{allparity} mode, we include contributions from odd-parity spectra, \textit{i.e.} those with odd $\sum_i \ell_i$. Assuming the Universe to be parity symmetric, these are unphysical. In \textsc{disconnected} mode, we estimate the Gaussian contribution to the 4PCF; this allows direct estimation of the purely non-Gaussian connected 4PCF, and will be discussed in \citet{4pcf_boss}.} 
    \label{tab: encore-modes}
\end{table}

\subsection{Main Algorithm}\label{subsec: algo-struc}

\subsubsection{Overview}\label{subsubsec: algo-overview}
The main process involved in computing NPCFs from a galaxy survey dataset is the estimation of multipoles, $\mathcal{N}^B_\L$ and $\mathcal{R}^B_\L$, from a discrete set of particle positions. As in \eqref{eq: NPCF-mult-discrete-binned-estimator}, this may be framed as a weighted pair-count, and an overview of the procedure is given in Algorithm \ref{algo: overview}. Essentially there are three processes of interest: (1) computing the $a_{\ell m}^b(\vs_i)$ harmonic coefficients at the location of each primary particle $i$, (2) accumulating the contributions to $\zeta_\L^B$, and (3) input/output operations, including loading the particles and coupling coefficients and saving the NPCF estimates. The latter is straightforward; the only complexity is in computation of the $C^\L_M$ coupling coefficients (defined in \ref{eq: basis-coupling}, with explicit forms given in \ref{eq: N=3-coeff},\,\ref{eq: N=4-coeff}\,\&\,\ref{eq: N=5-basis-and-coeff}). Since the determination of $C^\L_M$ requires Wigner 3-$j$ manipulations, they are precomputed in \textsc{python} using the \textsc{sympy} package\footnote{\href{https://www.sympy.org/en/index.html}{www.sympy.org}} and loaded at runtime. 

\begin{algorithm}
	\caption{Overview of the main algorithm underlying computation of the $N$-point correlation functions. Given precomputed coupling coefficients $C_M^\L$ \eqref{eq: basis-coupling}, this routine computes the correlation function estimates (with $N>2$) from $\Ng$ particles with angular momentum coefficients up to $\ell = \ell_\mathrm{max}$ using $N_r$ radial bins with minimum (maximum) separation $r_\mathrm{min}$ ($r_\mathrm{max}$). This is the main procedure underlying the \textsc{encore} code and implements the estimator of \eqref{eq: NPCF-mult-discrete-binned-estimator}. For windowed data, this approach is applied to two datasets: a field of random particles (which define the window) and one of `data-minus-randoms'. These are combination via \eqref{eq: edge-corrected-NPCF}.}
	\begin{algorithmic}[1]
	    \State Load particle positions and weights from disk
	    \State Read-in precomputed coupling coefficients $C_M^\L$ defined in \eqref{eq: basis-coupling}
		\For {primary particle $i=1,\ldots,\Ng$}
			\For {secondary particle $j=1,\ldots,\Ng$ with $j\neq i$}
			    \If{ $r_\mathrm{min}<|\vr_i-\vr_j|< r_\mathrm{max}$}
				    \For {$\ell\leq \ell_\mathrm{max}$}
				        \For{$m=0,\ldots,\ell$}
				            \For {$b=1,\ldots,N_r$ }
				                \State Accumulate the contribution to the harmonic coefficient $a_{\ell m}^b(\vs_i)$ using \eqref{eq: alm-binned-def} (cf.\,Algorithm\,\ref{algo: alm})
				            \EndFor
				        \EndFor
				    \EndFor
			    \EndIf
			\EndFor
			\State Compute the NPCF contribution as the product of $(N-1)$ harmonic coefficients at for each multiplet $\L$ and set of bin indices $B$, summed over $\{m_i\}$ (cf.\,\ref{eq: NPCF-mult-discrete-binned-estimator}\,\&\,Algorithm\,\ref{algo: summations}). 
		\EndFor
		\State Save the NPCF counts to disk
	\end{algorithmic}\label{algo: overview}
\end{algorithm}

Computing 
the $a_{\ell m}^b(\vs_i)$ harmonic coefficients at each primary particle location $\vs_i$ is less trivial. From \eqref{eq: binned-discrete-alm}, each can be evaluated as a spherical-harmonic-weighted sum of all secondary particles, $j$, whose separation from the primary lies within bin $b$. To implement this, we use the approach of \citet{2015MNRAS.454.4142S}, whereupon spherical harmonics are obtained from their Cartesian representations. For each secondary particle $j$, we first define the unit separation vector $\widehat{\vr_j-\vs_i}\equiv \hat{\vec \Delta} = (\Delta_x,\Delta_y,\Delta_z)$, and compute all (non-negative integer) powers $\Delta_x^p\Delta_y^q\Delta_z^r$ where $p+q+r\leq\ell_\mathrm{max}$. These can be combined to form the spherical harmonics. In practice, we compute histograms of $\Delta_x^p\Delta_y^q\Delta_z^r$ for all  required $j$ particles (skipping those with separations outside $[r_\mathrm{min},r_\mathrm{max}]$), before combining these to form the harmonic coefficients. This requires $\mathcal{O}(\Ng^2)$ operations and a total of $(1+\ell_\mathrm{max})(2+\ell_\mathrm{max})(3+\ell_\mathrm{max})/6$ histograms per bin $b$.

As an example, consider $a_{2-1}^b(\vs_i)$. This requires the spherical harmonic $Y_{2-1}(\hat{\vec \Delta}) \propto (\Delta_x-i\Delta_y)\Delta_z$; to compute it, we first obtain the sum of $\Delta_x\Delta_z$ and $\Delta_y\Delta_z$ over all secondary galaxies within the bin $b$, then combine the sums and normalize to obtain $a_{2-1}^b(\vs_i)$. \textit{Pseudo}-code for this procedure is shown in Algorithm\,\ref{algo: alm}. Due to the spherical harmonic conjugation symmetry, $a_{\ell-m}^b = a_{\ell m}^{b*}$, allowing us to compute only harmonic coefficients with $m\geq 0$.


\begin{algorithm}
    \caption{Procedure to calculate the $a_{\ell m}^b(\vs_i)$ harmonic coefficients \eqref{eq: alm-binned-def} for a given primary particle (either a galaxy or random) at position $\vs_i$. The algorithm relies on two facts: (1) $a_{\ell m}^b(\vs_i)$ is a sum over spherical harmonics $Y_{\ell m}(\widehat{\vr_j-\vs_i})$ for secondary $\vr_j$ within bin $b$, and (2) the spherical harmonics can be written as a sum of products of their Cartesian components. In the first part of the algorithm, we accumulate histograms of all relevant unit powers (here denoted $f_{pqr}$), before transforming these into spherical harmonics. In practice, the first loop is implemented using custom Advanced Vector Extension (AVX) instruction sets \citep{2015MNRAS.454.4142S}, and operates on sets of eight secondary particles simultaneously.} 
    \begin{algorithmic}[1]
        \For {secondary particle $j=1, \ldots,\Ng$}
            \State Compute the unit separation vector $(\Delta_x,\Delta_y,\Delta_z) = (\vs_i-\vr_j)/|\vs_i-\vr_j|$ 
            \State Find the radial bin $b$ corresponding to separation $|\vs_i-\vr_j|$
            \State Accumulate all combinations $f^b_{pqr}\pluseq w_j\Delta_x^p\Delta_y^q\Delta_z^r$ with $p+q+r\leq\ell_\mathrm{max}$ for secondary weight $w_j$
        \EndFor
        \For {$\ell=0,\ldots,\ell_\mathrm{max}$}
            \For {$m=0,\ldots,\ell$}
                \State Construct $a_{\ell m}^{b}(\vs_i)$ as a weighted sum of $f^b_{pqr}$ coefficients
        \EndFor
   \EndFor
    \end{algorithmic}\label{algo: alm}
\end{algorithm}

Finally, we must combine the $a_{\ell m}^b(\vs_i)$ harmonic coefficients to form the NPCF contributions. This is done using \eqref{eq: NPCF-mult-discrete-binned-estimator}, and must be performed $\Ng$ times. In essence, this is just an outer product of $(N-1)$ harmonic coefficients, summed over the $m_i$ indices and weighted by the coupling matrix $\mathcal{E}(\L)\;C^\L_M$. For large $N$, evaluation of this expression becomes expensive, since the dimensionality of $\L$, $M$ and $B$ is large (cf.\,\S\ref{subsec: estimator-discrete}). For each of element in $\zeta_\L^B$ (of which there are at most $\binom{N_r}{N-1}(1+\ell_\mathrm{max})^{N-1}(1+2\ell_\mathrm{max})^{N-4}$ for $N>3$), we must sum over a maximum of $(1+\ell_\mathrm{max})^{N-2}$ $M$ combinations, requiring a large number of operations to be performed for each primary particle.

In practice, this may be expedited. 
Firstly, we omit any $\Lambda$ contributions violating the triangle conditions, (e.g., for $N>3$, we can enforce $|\ell_1-\ell_2|\leq \ell_{12}\leq \ell_1+\ell_2$), and note that the $C_M^\Lambda$ conditions require 
$m_{N-1} = -m_1-\ldots -m_{N-2}$, as well as $|m_{12}|\leq \ell_{12}$, $|m_{123}|\leq\ell_{123}$ \textit{et cetera}. Usually, we compute only correlators obeying $\zeta(\hR) = \zeta(-\hR)$, \textit{i.e.} those with $\mathcal{E}(\L) = 1$. These correspond to even $\sum_i \ell_i$. Finally, we can use the conjugate symmetry of the harmonic coefficients to sum only terms with $m_{N-1}\geq 0$. This is detailed in Appendix \ref{appen: spin-sum}, and corresponds to rewriting the single-particle contribution to \eqref{eq: NPCF-mult-discrete-binned-estimator} as
\beq\label{eq: simple-NPCF-summation}
    w_i\sum_{m_1\ldots m_{N-1}}\mathcal{E}(\L)\;C_M^\Lambda a_{\ell_1m_1}^{b_1}(\vs_1)\ldots a_{\ell_{N-1}m_{N-1}}^{b_{N-1}}(\vs_i) &=& 2w_i\sum_{m_1+\cdots +m_{N-2}\leq 0}S(m_{N-1})\mathcal{E}(\L)\;C_M^\Lambda \,\mathbb{Q}_\L\left[a_{\ell_1m_1}^{b_1}(\vs_i)\cdots a_{\ell_{N-1}m_{N-1}}^{b_{N-1}}(\vs_i)\right],
\eeq
where $S(m) = 1/2$ if $m=0$ and unity else, and the operator $\mathbb{Q}_\L$ returns the real-part if $\L$ is parity-even and $i$ multiplied by the imaginary-part else. This reduces the number of terms that must be summed approximately by a factor of two, and makes manifest the reality of $\zeta_\L$ for even-parity multiplets. Algorithm \ref{algo: summations} gives an example of the full summation strategy for the 4PCF.

\begin{algorithm}
	\caption{Typical example of the summations required to compute the NPCF estimator, given the $a_{\ell m}^b$ harmonic coefficients for radial bins $b$ and angular momentum indices
	$\ell, m$. Here, we give the \textit{psuedo}-code for the parity-even $N=4$ summations; higher orders proceed similarly. The 4PCF is parametrized by three angular momenta, $\L = \{\ell_1,\ell_2,\ell_3\}$ (obeying the conditions $|\ell_1-\ell_2|\leq \ell_3\leq \ell_1+\ell_2$ and $\ell_i\leq \ell_\mathrm{max}$) and three radial bins $B=\{b_1,b_2,b_3\}$. Note that we store only bins with $b_1<b_2<b_3$; any other bins are either superfluous or contain zero-lag contributions. Here, we compute only even-parity multiplets, skipping any triplet with odd $\ell_1+\ell_2+\ell_3$, as indicated by the `continue' statement. We further note that `load' indicates that a quantity held in RAM is accessed and passed to the cache. $S(m)$ is defined below \eqref{eq: simple-NPCF-summation}.
	}
	\begin{algorithmic}[1]
		\For {$\ell_1=0,\ldots,\ell_\mathrm{max}$}
		    \For {$\ell_2=0,\ldots,\ell_\mathrm{max}$}
		        \For {$\ell_3=|\ell_1-\ell_2|,\ldots,\mathrm{min}\left(\ell_1+\ell_2,\ell_\mathrm{max}\right)$}
		        \State \textbf{if} $\ell_1+\ell_2+\ell_3$ is odd \textbf{continue} 
		            \For {$m_1=-\ell_1,\ldots,\ell_1$}
		                \For {$m_2=-\ell_2,\ldots,\ell_2$}
		                \State $m_3 = -m_1-m_2$
		                \State \textbf{if} $m_3>\ell_3$ \textbf{or} $m_3<0$ \textbf{continue}
		                \State Load coupling weight $w_i(-1)^{\ell_1+\ell_2+\ell_3}C_{m_1m_2m_3}^{\ell_1\ell_2\ell_3}$
		                \For {$b_1=1,\ldots ,N_r$}
		                    \State Load $a_{\ell_1m_1}^{b_1}$ or, if $m_1<0$, $(-1)^{m_1}a_{\ell_1-m_1}^{b_1,*}$ 
		                    \For {$b_2=b_1+1,\ldots,N_r$}
		                        \State Load $a_{\ell_2m_2}^{b_2}$ or, if $m_2<0$, $(-1)^{m_2}a_{\ell_2-m_2}^{b_2,*}$
		                        \For {$b_3=b_2+1,\ldots,N_r$}
		                            \State Load $a_{\ell_3m_3}^{b_3}$
		                            \State Accumulate 4PCF; $\zeta_{\ell_1\ell_2\ell_3}^{b_1b_2b_3}\pluseq2S(m_3)w_i(-1)^{\ell_1+\ell_2+\ell_3}C_{m_1m_2m_3}^{\ell_1\ell_2\ell_3}\mathrm{Re}\left[a_{\ell_1m_1}^{b_1}a_{\ell_2m_2}^{b_2}a_{\ell_3m_3}^{b_3}\right]$.
		                       \EndFor
		                      \EndFor
		                    \EndFor
		                \EndFor
		            \EndFor
		        \EndFor
		    \EndFor
		\EndFor
	\end{algorithmic}\label{algo: summations}
\end{algorithm} 

\subsubsection{CPU Implementation}\label{subsubsec: cpu}
\textsc{encore} provides a C\textsc{++} implementation of the above routine. Whilst most of the code is straightforward, there are several points of note. Firstly, when computing the harmonic coefficients for each primary particle, we optionally employ custom Advanced Vector Extension (AVX) instruction sets to accumulate the necessary unit vector powers $\Delta_x^p\Delta_y^q\Delta_z^r$ (cf.\,\S\ref{subsubsec: algo-overview}). These were developed for the \textsc{Abacus} simulation project \citep{metchnik09,2019MNRAS.485.3370G,abacus21} 
and used in the C\textsc{++} implementation of the isotropic 3PCF algorithm of \citet{2015MNRAS.454.4142S}. Practically, this allows optimal vectorization, with eight secondary particles being analyzed simultaneously (assuming double precision). Whilst the $a_{\ell m}^b$ computation is not rate-limiting at large $N$, the addition of AVX instruction sets gives significant speed-boosts for the 3PCF (cf.\,\S\ref{subsec: scalings}).

Secondly, the code is (optionally) parallelized using \textsc{OpenMP}. For this, we distribute the iteration over $\Ng$ primary particles to as many threads as are available, parallelizing steps three to twelve of Algorithm \ref{algo: overview}. Each thread has its own value of the NPCF sum and weight functions, with the former then being combined at the conclusion of the algorithm. Only the list of particle positions is shared between threads. As shown in \S\ref{subsec: scalings}, the computation time scales as $1/N_\mathrm{threads}$ for the higher-point functions, implying that the parallelization is almost perfect. 

\subsubsection{GPU Implementation}\label{subsubsec: gpu}
Given that the bulk of the NPCF algorithm requires simple additions and multiplications of harmonic coefficients, we expect that the performance can be significantly enhanced using Graphical Processing Units (GPUs). Modern GPUs contain thousands of individual processing cores, each of which is capable of handling $\sim$\,$1000$ threads concurrently. Developers can take advantage of this by using platforms such as \textsc{nvidia}'s Compute Unified Device Architecture (\textsc{cuda}) to design massively parallel algorithms, resulting in significant speed gains for algorithms that are limited by computational resources rather than memory access. 
The \textsc{encore} code includes GPU acceleration via \textsc{cuda}, and it is found to be of significant use for the 5PCF and beyond (\S\ref{subsec: gpu-testing}). Since the computation time of the algorithm is a strongly increasing function of $N$ (due to the larger dimensionality of $\L$ and $B$), we focus on the step which is rate-limiting for the higher-point functions; combination of harmonic coefficients to form the NPCF summand. \textit{i.e.} step 11 in Algorithm \ref{algo: overview}. Currently, we do not include a GPU implementation of the $a_{\ell m}^b$ estimation; this may be added in the future. We note also that the GPU code is not currently compatible with \textsc{OpenMP}, \textit{i.e.} it should be run single-threaded.

To implement a GPU-based code, we must first consider what functions we wish each GPU thread to compute. Two options come to mind: (a) allow each thread to compute the $\zeta_\L^B$ contribution from a single multiplet $\L$, set of bin-indices $B$, summing over all choices of $M$, or (b) use a separate thread for each $M$ element. Option (b) allows a greater number of threads to be utilized, maximizing the GPU capabilities; further, each thread must perform the same number of operations. 
However, since multiple $M$ contribute to the same multiplet $\L$, this requires careful consideration of memory access. \textsc{cuda} requires thread blocks to be able to execute independently, in any order, and with many threads running simultaneously it is possible for a collision to occur during which two threads attempt to write to the same NPCF element at once, resulting in an undefined value.  In order to avoid this error, the addition into the NPCF array $\zeta_\L^B$ must be done atomically; \textit{i.e.} only one thread can access a given NPCF element at once. This will lead to reduced performance. For option (a), each thread computes a different element of $\zeta_\L^B$ thus there are no issues with atomicity, though the total number of threads utilized (equal to the dimension of the NPCF) is smaller. This approach does not achieve maximum parallelization if the dimension is small and the work per thread varies as a function of $\L$.

In practice, we implement both approaches within \textsc{encore} for the 4PCF and 5PCF, and compare their performance in \S\ref{subsec: gpu-testing}. Schematically, the action of the GPU-parallelized code is the following:
\begin{itemize}
    \item For each primary position $i$, compute $a_{\ell m}^b$ on the CPU.
    \item Copy the $a_{\ell m}^b$ array to the GPU.
    \item For each multiplet $\L$ and set of bin-indices $B$, compute the contribution to $\zeta$ on the GPU.
    \item After all primary particles have been processed, 
    copy the result back to the CPU.
\end{itemize}
The GPU kernels are launched a total of $\Ng$ times. Due to the conditions on $\L$, $B$ and $M$, it is non-trivial to relate the GPU thread index to the individual angular momentum and bin components; for this reason we first generate a look-up table for $\L$, $B$ and (optionally) $M$, which is passed to the GPU on initialization. This additionally avoids any threads having to do only trivial work; for example we can ensure that any combinations that violate the triangle conditions are not included. Finally, we note that some commerically available GPUs are optimized for float-based calculations whilst others perform well using doubles. For this reason we consider three variants: (1) using double precision at all times (as for the CPU calculation), (2) using single precision only, and (3), using single precision for the $a_{\ell m}^b$ arrays, but adding to the NPCF as a double. The performance of each variant is compared in \S\ref{subsec: gpu-testing}.

\subsection{Reconstructing the \resub{Geometry-Corrected} NPCF}\label{subsec: edge-correction-in-practice}
\S\ref{subsec: algo-struc} presented a detailed discussion of how to compute the NPCF multiplets for a given set of input particles. We now consider the practicalities of computing the \textit{full} geometry-corrected NPCF, as outlined in \S\ref{subsec: edge-correction}.

For this purpose, we require two fields: the data (galaxy positions), $D(\vr)$, and randoms (unclustered particles), $R(\vr)$ (cf.\,\ref{eq: data-randoms-def}). The latter are used to compute the normalization term $\mathcal{R}_\L^B$ appearing in \eqref{eq: edge-corrected-NPCF}; this requires a straightforward application of estimator \eqref{eq: NPCF-mult-discrete-binned-estimator}. For the second term, $\mathcal{N}_\L^B$, we first compute a joint `data-minus-randoms' field by combining the data and random catalogs together (following \citealt{2015MNRAS.454.4142S}). Randoms are assigned negative weights and normalized such that the summed weight is zero. The multiplets of this field give $\mathcal{N}_\L$ directly.
\footnote{An additional method to compute $\mathcal{N}_\L^B$ is to expand the numerator of \eqref{eq: generalized-landy-szalay} into separate terms, each involving just data or randoms. Schematically, one writes
\beq
    \mathcal{N}_\L^B \equiv [(D-R)^N]_\L^B = \left[D^N\right]_\L^B - N\left[D^{N-1}R\right]_\L^B + \cdots  - N\left[DR^{N-1}\right]_\L^B + \left[R^N\right]_\L^B,
\eeq
where $\left[X_1X_2\ldots X_N\right]_\L^B$ are the multiplets of $N$ input fields $X_1, X_2 \ldots , X_N$. This is analogous to the standard procedure for the 2PCF computation (using $DD$, $DR$, and $RR$ counts), but somewhat inefficient here, since it requires significant recomputation.}


In practice, the random catalog $R(\vr)$ contains many more particles than the data to minimize the inherent Poissonian errors. \citet{2015MNRAS.454.4142S} and \citet{2019A&A...631A..73K} advocate splitting the random catalog into subsets, each of size $\sim$\,$1.5-2\times$ that of the data, and co-adding the $\mathcal{N}_\L^B$ counts for each; if the computation time is quadratic in the number of particles, this minimizes the work required for a given level of Poisson noise.\footnote{For periodic-box geometries, it is possible to partially remove the dependence on random catalogs, as discussed in \citet{2019MNRAS.486L.105P}, \citet{2020JCAP...12..021S} and \citet{2021MNRAS.501.4004P}.} Whilst appropriate for the 3PCF, the runtime of \textsc{encore} when applied to higher-point statistics is usually found to scale linearly with the number of particles  (cf.\,\S\ref{subsec: scalings}), thus there is no gain to such a split. It is retained in \textsc{encore} for full generality however, leading to $\sim$\,$32$ separate catalogs if one uses randoms with $50\times$ the galaxy density. Given that the $a_{\ell m}^b$ contributions from the data must be included in each split, the approach is somewhat inefficient; to avoid this, we include the functionality to save and reload the histogrammed Cartesian powers of positively-weighted (data) particles. Finally, we note that, since the $\mathcal{R}_\L^B$ counts use only positively weighted particles, their intrinsic noise is much reduced, thus it is sufficient to use only a single random partition for their computation. 

Given the estimated $\mathcal{R}_\L^B$ and 
$\mathcal{N}_\L^B$ quantities (averaged over the random subsets), we can compute the full edge-corrected NPCF via \eqref{eq: edge-corrected-NPCF}. To form the coupling matrix $M_{\L\L'}$ we require the generalized Gaunt integrals of \eqref{eq: generalized-gaunt-def}; these involve Wigner 9-$j$ symbols and can be straightforwardly precomputed for a given $N$ and $\ell_\mathrm{max}$ using the \textsc{python} \textsc{sympy} package. For simplicity, the entire edge-correction procedure is performed in \textsc{python}, and involves a matrix inversion for each radial bin combination $B$. In the case of a periodic box geometry, the edge-correction is trivial, and we use the analytic $\mathcal{R}_\L^B$ functions of \eqref{eq: R-periodic-form}. To facilitate general use of the \textsc{encore} code, we provide a \textsc{bash} (or \textsc{slurm}) script to automate the entire process of computing raw NPCF multiplets and applying edge-correction, allowing for the full NPCF estimates to be obtained given the data and random particle positions.

\section{Results}\label{sec: results}
We now present results from applying the NPCF estimators of \S\ref{sec: estimator} to data, considering both scalings with computational resources and particle number density, as well as a demonstration of the edge-corrected 4PCF and 5PCF of realistic simulations. 

\subsection{\resub{Dimensionality and Scalings}}\label{subsec: scalings}

\resub{Before presenting numerical results, we consider the dimension of the $\zeta_\L^B$ NPCF statistic, which will facilitate later discussion of computational scalings.} Assuming $N_r$ radial bins per dimension, and asserting that radial bins do not overlap (cf.\,\S\ref{subsec: estimator-discrete}), $B$ contains $\binom{N_r}{N-1}$ elements, which is \resub{asymptotically equal to} $N_r^{N-1}$. The NPCF requires isotropic basis functions of $(N-1)$ coordinates; for $N>3$ these contain $(N-1)$ principal angular momenta $\ell_i$, each of which runs from $0$ to $\ell_\mathrm{max}$, and $(N-4)$ intermediate angular momenta $\ell_{12\ldots}$, each of which runs from $0$ to $2\ell_\mathrm{max}$ (cf.\,\S\ref{subsec: basis-def}). This implies that the number of elements in $\L$ is bounded from above by $(1+\ell_\mathrm{max})^{N-1}(1+2\ell_\mathrm{max})^{N-4}$; the true number is somewhat smaller due to the triangle conditions. Furthermore, we must sum over $(N-2)$ integers $m_i$, each of which can run from $-\ell_i$ to $\ell_i$ (with $m_{N-1}$ set by $\sum_i m_i = 0$), giving a maximum of $(1+2\ell_\mathrm{max})^{N-2}$ summation terms in \eqref{eq: NPCF-mult-binned-estimator} for each choice of $\L$ and $B$. This further bounds the dimensionality of the coupling coefficient $C^\L_M$ to $(1+\ell_\mathrm{max})^{N-1}(1+2\ell_\mathrm{max})^{2N-6}$. Asymptotically, the dimension of the NPCF statistic, $\zeta_\L^B$, scales as $N_r^{N-1}\ell_\mathrm{max}^{2N-5}$, which quickly becomes very large as $N$ increases. Given that our eventual goal is to compute $\mathcal{O}(10)$ parameters from the statistic, it is likely that data compression methods \citep[e.g.,][]{2000ApJ...544..597S,2021PhRvD.103d3508P} will prove useful in any higher-point NPCF analysis.

The algorithm described in \S\ref{sec: implementation} contains two computationally intensive steps. The \resub{runtimes of these} have the following scalings:
\begin{itemize}
    \item \textbf{Computation of $a_{\ell m}^b(\vs_i)$}: This requires a count over all pairs of galaxies separated by distances within a spherical shell $[r_\mathrm{min},r_\mathrm{max}]$; as such, the work scales as $\Ng\times \bar n\left(r_\mathrm{max}^3-r_\mathrm{min}^3\right)$ for (galaxy or random) number density $\bar n$ and total number $\Ng$, \textit{i.e.} it is quadratic in the total number of particles if $\bar n$ is fixed. Computation of each count requires accumulation of $(1+\ell_\mathrm{max})(2+\ell_\mathrm{max})(3+\ell_\mathrm{max})/6$ powers of $\Delta_x^p\Delta_y^q\Delta_z^r$ (cf.\,\S\ref{subsec: algo-struc}), 
    thus runtime is cubic in $\ell_\mathrm{max}$ and linear in $N_r$. Note that this is independent of the NPCF order, $N$.
    \item \textbf{Accumulation of $\zeta_\L^B$}: For each $\L$ and $B$, we must sum over all $M$ indices for $(N-1)$ harmonic coefficients. Since this must be performed for each primary position, computation time scales as $\Ng \equiv \bar nV$. Restricting to bins with $b_1<b_2<\ldots<b_{N-1}$, there are $\binom{N_r}{N-1}$ combinations contained in $B$, with $(N-2)$ independent $M$ elements for each of $(2N-5)$ choices for $\L$ \resub{(as discussed above)}. 
    This implies the asymptotic scaling is of the form $\ell_\mathrm{max}^{3N-7}N_r^{N-1}$, which is a strong function of $\ell_\mathrm{max}$.
\end{itemize}
Given this, we expect the runtime of \textsc{encore} to scale as $\Ng^2$ for dense samples at low $N$ (as found in \citealt{2017arXiv170900086F}), or $\Ng$ for higher-point functions with fewer particles, or those with a large number of bins.

To test the above, we compute the parity-even NPCF for a synthetic box of galaxies at different mean densities $\bar n$. We use a cubic box with $L_\mathrm{box} = 1500\Mpch$, populated with randomly placed galaxies of varying number density from one-tenth to ten times that of BOSS CMASS (\citealt{2015ApJS..219...12A}, cf.\,\S\ref{subsec: mocks}). The pair counts are computed on a 16-core machine in \textsc{periodic} mode (cf.\,Tab.\,\ref{tab: encore-modes}), using $\ell_\mathrm{max} \in \{10,10,5,3\}$ for $N \in \{3,4,5,6\}$ respectively. The maximum multipole is varied between samples to keep the dimensionality of the statistics reasonable, but the number of radial bins is fixed to $N_r = 10$; here, the dimensionality of $B$ and $\L$ is $\{46, 123, 215, 259\}$ and $\{6, 381, 1211, 1558\}$ respectively, incorporating all triangle and binning conditions. The total dimension is the product of these, \textit{i.e.} approximately $\{280,4.7\times 10^4, 2.6\times 10^5, 4.0\times 10^5\}$ elements respectively. 
The run-times, $T$, of \textsc{encore} for each setup are shown in Fig.\,\ref{fig: n-scaling}.

\begin{figure}
    \centering
    \includegraphics[width=0.6\textwidth]{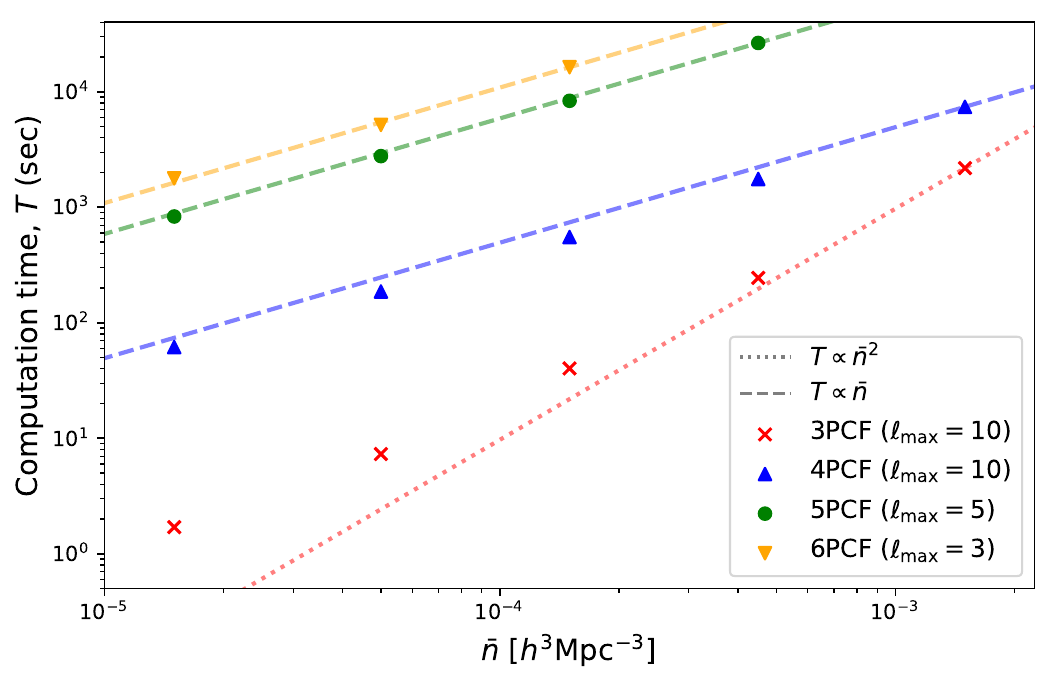}
    \caption{Scaling of the \textsc{encore} code with the mean number density of the sample, when run in 3PCF, 4PCF, 5PCF and 6PCF mode with $10$ radial bins. As described in \S\ref{subsec: scalings}, the galaxy sample is simply a set of uniformly distributed points in a periodic box. The scaling is close to quadratic (linear) with number density and hence particle number $\Ng$ for the 3PCF (4PCF and above), as expected. The dotted (dashed) lines show quadratic (linear) relationships, anchored through the highest number-density sample. Note that the maximum multipole, $\ell_\mathrm{max}$ is varied to keep the dimensionality of the output reasonable; its value is indicated by the caption in each case. All calculations are performed on sixteen $2.4$\,GHz Intel Skylake CPUs, parallelized using \textsc{OpenMP}. We show only results computable within 16 hours of wall-clock time.}
    \label{fig: n-scaling}
\end{figure}

Notably, the four and higher-point algorithms follow closely a $T$$\sim$$\Ng$ scaling (noting that $\Ng\propto \bar{n}$, since $V$ is fixed). The 3PCF exhibits a quadratic scaling $\Ng^2$ at high $\bar{n}$, though is closer to linear for low density samples. Given the above discussion on the limiting steps of each algorithm, this is as expected. Furthermore, the low-$\bar{n}$ plateau of the 3PCF is due to other processes from Algorithm \ref{algo: overview} that become important, such as the loading of galaxies and coupling weights into memory and the particle neighbor searches. At much larger $\bar{n}$, we may expect the 4PCF to scale as $\Ng^2$, as it becomes limited by the same processes as for the 3PCF. Due to the prohibitive $N_r^{N-1}\ell_\mathrm{max}^{3N-7}$ scaling of the $\zeta_\L^B$ accumulation step, the quadratic scaling is unlikely to occur for higher-point functions at any reasonable $\bar{n}$.

It is also useful to quantify how well the \textsc{encore} algorithm can be parallelized. To this end, we rerun the analysis used to create Fig.\,\ref{fig: n-scaling} at $\bar{n}\approx1.5\times 10^{-4}h^3\mathrm{Mpc}^{-3}$, but on a varying number of CPU threads. The results are shown in Fig.\,\ref{fig: cpu-scaling} and the conclusion is straightforward; the algorithm exhibits almost perfect scaling with the number of cores. This is as expected; as discussed in \S\ref{subsubsec: cpu}, the \textsc{OpenMP} implementation parallelizes the sum over all primary particles, each of which is independent (and holds its own copy of the NPCF sum), except for look-ups in the shared particle grid. We note that the code does not currently include Message Passing Interface (MPI) parallelization; whilst this could lead to further speed boosts and distribute the work across multiple nodes, it is expected that a more useful application of these resources would be to compute separate $\mathcal{N}_\L$ counts (cf.\,\ref{eq: edge-corrected-NPCF}), or the NPCF estimates of different simulations. Fig.\,\ref{fig: cpu-scaling} may also be used to examine the dependence of the runtime on $\ell_\mathrm{max}$. The ratio of $\ell_\mathrm{max} = 10$ to $\ell_\mathrm{max} = 5$ takes the values $1.9\pm 0.1$, $11.5\pm 1.1$ for the 3PCF and 4PCF; this is somewhat better than the scalings discussed above, indicating the asymptotic limit (\textit{i.e.} the scalings for $\ell_\mathrm{max}\gg 1$) 
has not yet been reached.

\begin{figure}
    \centering
    \includegraphics[width=0.6\textwidth]{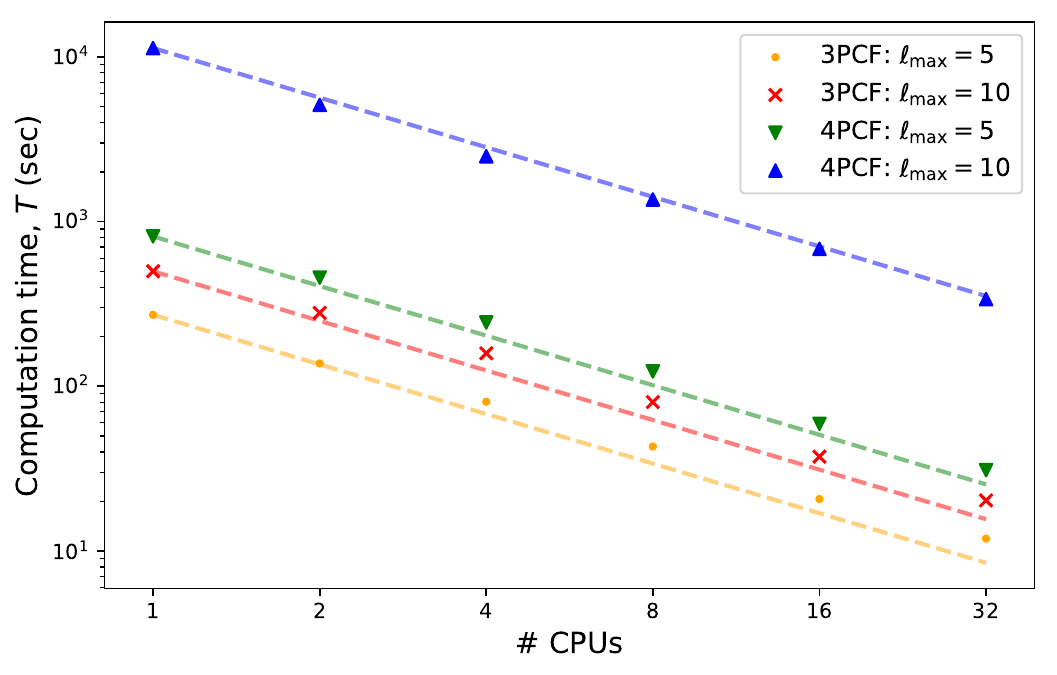}
    \caption{Strong scaling of the \textsc{encore} code: dependence of runtime, $T$, on the number of CPU cores on a single node, for the test cases considered in Fig.\,\ref{fig: n-scaling}, fixing $\bar{n} \approx 1.5\times 10^{-4}$. Dashed lines indicate linear relationships, and are calibrated at the single-CPU time. Results are shown for $N=3$ and $N=4$ with two choices of $\ell_\mathrm{max}$. The code exhibits close-to-ideal parallelization, as shown by the linear trends. This extends also to higher order. As expected, the runtime increases significantly when $\ell_\mathrm{max}$, and thus the number of multiplets $\L$, is increased. The 3PCF results deviate slightly from the linear relationship when using 32 CPUs; this occurs since we become limited by other processes such as data loading.}
    \label{fig: cpu-scaling}
\end{figure}

\subsection{Comparison to Brute-Force NPCF Estimators}
Given the complexity of the \textsc{encore} algorithm (\S\ref{sec: estimator}\,\&\,\S\ref{sec: implementation}), it is important to test whether the code returns accurate estimates of the $N$-point functions. For this purpose, we compared the NPCF multiplets output by \textsc{encore} to those computed via a brute-force approach. To form such estimates, we compute $\zeta_\L^B$ directly, as a sum over $\Ng$ particles weighted by the relevant basis function:
\beq\label{eq: brute-force-npcf}
    \zeta_\L^B = \sum_{i_1\ldots i_N}\P_\L^*(\widehat{\vr_{i_2}-\vr_{i_1}},\ldots,\widehat{\vr_{i_N}-\vr_{i_1}})\Theta^{b_1}(|\vr_{i_2}-\vr_{i_1}|)\cdots\Theta^{b_N}(|\vr_{i_1}-\vr_{i_N}|),
\eeq
where $\vr_i$ is the position vector of particle $i$, and $\Theta^b(r)$ are binary binning functions, as in \S\ref{subsec: estimator-continuous}. This is simply the definition for the NPCF multiplets \eqref{eq: extract-npcf-mult} for discrete data, with the addition of binning. Since it involves a sum over $N$ particles, it has complexity $\mathcal{O}(\Ng^N)$. To evaluate the angular basis functions, we use the Cartesian forms given in \citet[Appendix A]{2020arXiv201014418C}; for example, $\P_{110}(\hr_1,\hr_2,\hr_3) = -\sqrt{3}(4\pi)^{-3/2}(\hr_1\cdot\hr_2)$. This approach is prohibitively slow for large $\Ng$, thus we use only $\Ng = 20$ as a test case. In our case, the two approaches agreed to machine precision for all $\L$ considered and $N$ in $\{3,4,5,6\}$. Since this approach knows nothing about spherical harmonics or angular momenta, the precise agreement affords us confidence in the validity of \textsc{encore}.

\subsection{GPU Testing}\label{subsec: gpu-testing}
We now consider the test \textsc{encore}'s GPU implementation. As in \S\ref{subsubsec: gpu}, GPUs are currently used to expedite the accumulation of the $\zeta_\L^B$ arrays for each primary particle, given $a_{\ell m}^b(\vs_i)$ coefficients computed on the CPU. Two approaches are possible: (a) each GPU thread can sum over all $M$ for a given $\L$ and $B$, or (b) each GPU thread can compute contributions from a single set of coefficients $M$ (requiring atomic memory access). To test these, we use the same setup as \S\ref{subsec: scalings}, except with $\bar{n}$ fixed to the lowest value of $1.5\times 10^{-5}h^{3}\mathrm{Mpc}^{-3}$ for speed (recalling that the higher-point functions scale linearly with $\bar{n}$). The 4PCF and 5PCF are then computed both on the CPU and GPU, using an Nvidia P100 chip for the latter, and the runtimes are shown in Tab.\,\ref{tab: gpu-testing}.

For the 4PCF at $\ell_\mathrm{max} = 5$, computation is far faster for the 16-core CPU code than for the GPU-accelerated version. This occurs since the GPU code uses only a single CPU to compute the $a_{\ell m}^{B}$ harmonic coefficients, which become rate-limiting for low $N$ and $\ell_\mathrm{max}$. Testing the GPU implementation against such a parallelized CPU code is a reasonable test however, given that GPUs are currently a scarcer resource for most high performance computing systems, though this is subject to change. 
For $\ell_\mathrm{max} = 10$, we find a slight preference for the GPU code, which is much increased if one considers the 5PCF. Indeed, the GPU implementation is able to provide speed-ups of $\mathcal{O}(8\times)$ for the latter case. 

On newer GPU chips such as the A100 series,\footnote{\href{https://www.nvidia.com/en-us/data-center/ampere-architecture/}{www.nvidia.com/en-us/data-center/ampere-architecture/}} we expect the performance to increase significantly. This is possible since the primary particle loops entering Algorithm \ref{algo: summations} are distributed across a large number of threads. 
For the P100 chip used herein, we have 3584 \textsc{cuda} cores, each with up to 2048 threads, allowing a maximum of $\sim$\,$7.3\times10^6$ threads to be run in parallel. The A100 chip has 6912 CUDA cores and moreover, as each individual core is faster, Nvidia claims an order of magnitude speed-up for HPC applications.\footnote{\href{https://www.nvidia.com/en-us/data-center/a100/}{www.nvidia.com/en-us/data-center/a100/}}

\begin{table}
    \centering
    \begin{tabular}{l|c|c|c}
    \textbf{Mode} &  \textbf{Wall-clock time} (4PCF, $\ell_\mathrm{max}=5$) & \textbf{Wall-clock time} (4PCF, $\ell_\mathrm{max}=10$) & \textbf{Wall-clock time} (5PCF, $\ell_\mathrm{max}=5$)\\\hline\hline
    CPU (16-core) & $4.8$ & $58.8$  & $741.8$ \\\hline
    GPU-a (double) & $31.1$ & $41.5$ & $86.1$ \\
    GPU-a (float) & $21.9$ & $38.4$ & $74.3$ \\
    GPU-a (mixed) & $22.0$ & $38.6$ & $77.2$ \\\hline
    GPU-b (double) & $30.1$ & $43.3$ & $177.7$ \\
    GPU-b (float) & $23.9$ & $42.7$ & $202.8$ \\
    GPU-b (mixed) & $30.6$ & $41.2$ & $159.6$ \\
    \end{tabular}
    \caption{Runtimes of the \textsc{encore} GPU code for the 4PCF and 5PCF using $N_r = 10$ radial bins and $\ell_\mathrm{max} = 5$ or $10$. In each case, we use a synthetic dataset containing $\approx 5\times 10^4$ uniformly distributed particles in a periodic box of side-length $L = 1500\Mpch$ (corresponding to one-tenth of the BOSS CMASS density), and count pairs of particles up to $r_\mathrm{max} = 200\Mpch$. All computations are performed on a single P100 Nvidia GPU coupled with an Intel Xeon Broadwell CPU. We give results for each of the GPU modes detailed in \S\ref{subsubsec: gpu}; these vary whether the $M$ summation is distributed across threads and what level of precision is used for intermediate calculations. For comparison, we give also the timings from the CPU code using a 16-core Intel Skylake node, parallelized using \textsc{OpenMP}. All times are reported in seconds.}
    \label{tab: gpu-testing}
\end{table}

Tab.\,\ref{tab: gpu-testing} also allows us to compare the different GPU modes discussed in \S\ref{subsubsec: gpu}. For the 4PCF test-cases, we find comparable runtimes between cases (a) and (b). Case (a) uses a much smaller number of tasks of unequal length (\numprint{8280} or \numprint{45720} for $\ell_\mathrm{max} = 5$ or $10$), whilst case (b) launches a far greater number of tasks (\numprint{135120} or \numprint{2499480} respectively), but requires atomic memory access (to ensure that no two threads access the same memory location concurrently). Whilst the first case does not come close to fully utilizing the GPU resources (since the number of tasks launched is significantly below the number that can be simultaneously performed on the GPU), atomic addition does not have to be performed, compensating for the inefficiency.
In fact, the degree of similarity suggests that memory access of the NPCF array is the main limiting factor in this case. 
For the 5PCF, we note a different story; case (a) is $\sim$\,$2-3\times$ faster than case (b), even though the number of tasks is much reduced (\numprint{295,260} instead of \numprint{29606430}, cf.\,\S\ref{subsec: scalings}). This is unsurprising since the 5PCF array is of much higher dimension than the 4PCF, thus we utilize a good fraction of the available GPU threads in case (a), while in case (b) we vastly exceed maximum parallelization; \textit{i.e.}, the number of tasks is larger than the maximum number of threads that can be processed in parallel. Above this threshold, we cease to realize additional speed gains so the difference in number of parallel threads is not large enough to overcome the speed penalty of atomicity.

We finally comment on the dependence of computation-time on precision (cf.\,\S\ref{subsubsec: gpu}). Most consumer-grade GPUs are optimized for float-based calculations, thus one might na\"ively expect a significant performance increase when lower precision is used in the GPU kernels; indeed this was found in preliminary testing on commercially-available GeForce GTX 1660 architectures where `float' mode was found to be $\sim$\,$8\times$ faster than `double' mode. However, many research-grade data center GPUs are optimized for double-precision calculations. We observe this in practice, as we do not find a particularly large improvement in efficiency on the Nvidia P100 GPUs; at most the runtime decreases by $\sim$\,$30\%$, though this depends on the precise kernel. We caution that reducing the precision of the GPU calculation necessarily leads to errors in the final result; for the 4PCF in `float' (`mixed') mode, the average fractional error is $\sim$\,$3\times 10^{-3}$ ($1\times 10^{-5}$), increasing to $\sim$\,$2\times 10^{-2}$ ($7\times 10^{-5}$) for the 5PCF.


\subsection{Application to Mock Catalogs}\label{subsec: mocks}
Finally, we apply the NPCF estimators of this work to realistic simulations of galaxy surveys. Here, we use a set of $10$ \textsc{MultiDark-Patchy} mocks \citep{2016MNRAS.456.4156K,2016MNRAS.460.1173R} constructed for the Baryon Oscillation Spectroscopic Survey (BOSS) \citep{2013AJ....145...10D}, using the twelfth data release (DR12) of SDSS-III \citep{2011AJ....142...72E,2015ApJS..219...12A}. Presentation of the full BOSS 4PCFs will follow in \citet{4pcf_boss}. The \textsc{MultiDark-Patchy} mocks are constructed using approximate gravity solvers, calibrated to an $N$-body simulation \citep{2015MNRAS.447.3693K}. Here, we use the northern Galactic cap (NGC) CMASS sample (hereafter CMASS-N), containing $\sim$\,$6\times 10^5$ galaxies with mean density $\bar{n}\sim1.5\times10^{-4}h^{3}\mathrm{Mpc}^{-3}$ at an effective redshift of $z_\mathrm{eff} = 0.57$. To use the NPCF estimator of \eqref{eq: generalized-landy-szalay}, we require also the random field $R(\vr)$; for this we use a public random catalog, containing 50 times the galaxy density. As in \S\ref{subsec: edge-correction-in-practice}, this is split into $32$ disjoint subsets, enhancing the computational efficiency for small $N$.

To give insight into the performance of the algorithm on real data, we first consider the runtime of the various sections of the NPCF computation in Algorithm \ref{algo: overview} for a single `data-minus-random' chunk. This is performed for both the 3PCF and 4PCF in 10 linearly-spaced radial bins up to $r_\mathrm{max} = 200\Mpch$, using $\ell_\mathrm{max} = 5$ or $10$ and skipping any odd-parity modes. This gives a total of 270 (495) correlation function components for the $\ell_\mathrm{max} = 5$ ($10$) 3PCF, and $13,320$ ($80,520$) for the 4PCF, as in \resub{\S\ref{subsec: scalings}}. The code is run on a 16-core node with a $2.4\,$GHz Intel Skylake processor (without GPU acceleration), and requires a wall-clock time of $36\,$s to compute both the 3PCF and 4PCF multipoles up to $\ell_\mathrm{max} = 5$, or $291\,$s with $\ell_\mathrm{max}=10$. 

Fig.\,\ref{fig: timings-breakdown} shows the timing breakdown for the above samples. We immediately note that the computation time is dominated by the two processes highlighted in \S\ref{subsubsec: algo-overview}: estimation of $a_{\ell m}^b(\vs_i)$ at each primary particle position and their combination (summing over $M$) to form the $\mathcal{N}^B_\L$ basis coefficients. For the 3PCF, the former is rate-limiting (as in \citealt{2015MNRAS.454.4142S}), whilst for the 4PCF, the bin-summations start to become important, particularly as $\ell_\mathrm{max}$ (and thus the number of $m$ indices) increases (cf.\,\S\ref{subsec: scalings}). Indeed, for moments beyond the 4PCF, virtually all computation time is spent in this procedure, thus the algorithm is expected to scale linearly with $\Ng$, as found in Fig.\,\ref{fig: n-scaling}. In practice, much of the time is spent loading the various harmonic coefficients elements into memory, rather than performing the actual addition into the NPCF sum itself. For $\ell_\mathrm{max} = 10$, and $N_r = 10$, we have $660$ complex elements in $a_{\ell m}^b(\vs_i)$, with a total size of $10.3$\,kB in double precision. This can comfortably fit within the $64$\,kB L1 cache of the Intel Skylake processors, which allows for the fastest memory access (though this is not true for the much larger NPCF array).

\begin{figure}
    \centering
    \includegraphics[width=\textwidth]{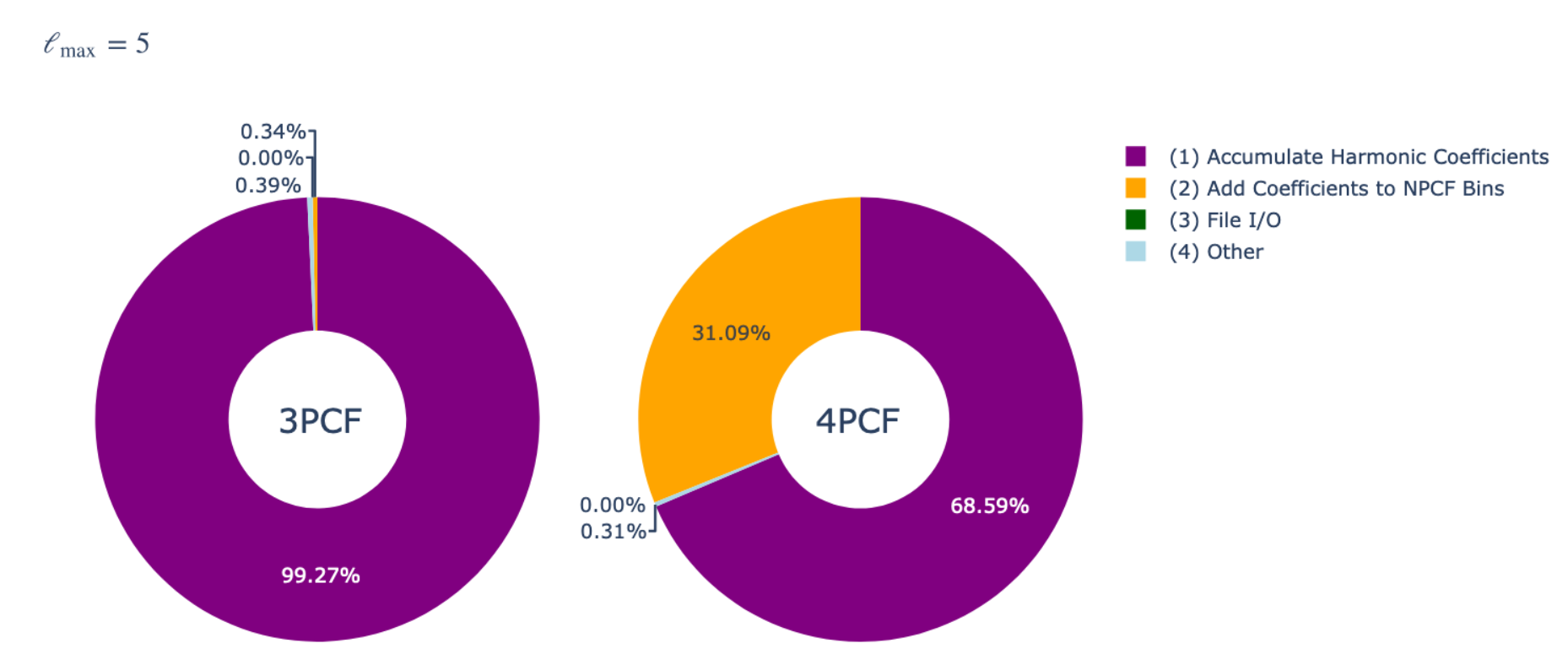}
    \includegraphics[width=\textwidth]{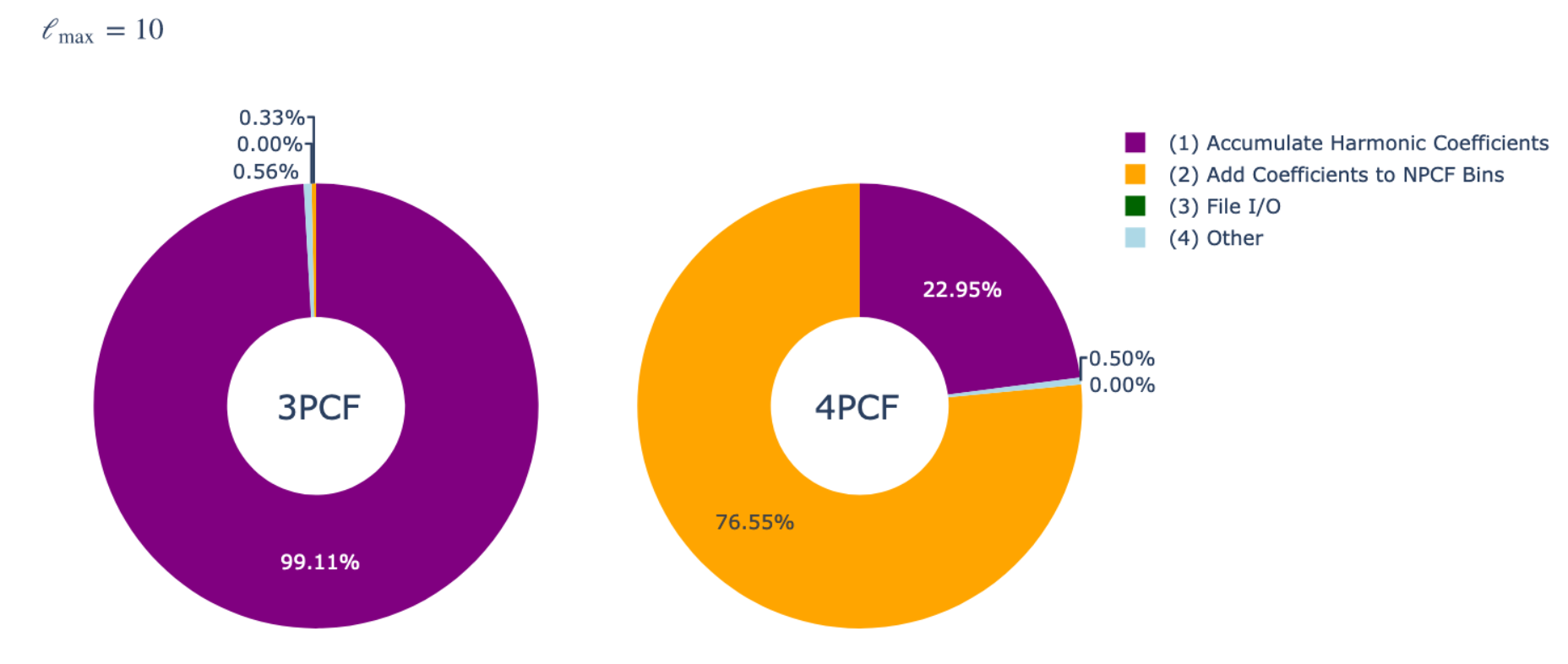}
    \caption{Breakdown of computation time for NPCF estimation using a single \textsc{MultiDark-Patchy} galaxy mock, for $N=3$ (left) and $N=4$ (right), with $\ell_\mathrm{max}=5$ (top) or 10 (bottom). In both cases, we use $10$ linearly spaced radial bins up to $r_\mathrm{max} = 200$, and run on a 16-core Intel Skylake processor. Runtime is broken down into four sections: (1) computation of the $a_{\ell m}^b$ harmonic coefficients around each primary particle, (2) combination of harmonic coefficients into NPCF multiplets, including summation over $M$ indices, (3) file input/output, (4) other, including placing particles onto a Cartesian grid and all other pre- and post-processing steps. Note that the runtime is dominated by the first two operations (which scale as $\Ng^2$ and $\Ng$ respectively, for particle number $\Ng$), with the summation over harmonic coefficients being the limiting factor at large $\ell_\mathrm{max}$ and high order. This motivates using GPUs to accelerate step (2) in these regimes.}
    \label{fig: timings-breakdown}
\end{figure}

Next, we consider estimation of the NPCFs themselves, including edge-correction. Here, we will present only the isotropic 4PCF and 5PCF, given that the 3PCF has been presented elsewhere \citep[e.g.,][]{2017MNRAS.469.1738S}. As before, we use $N_r = 10$, but restrict to $\ell_\mathrm{max} = 5$ ($\ell_\mathrm{max} = 3$) for the 4PCF (5PCF) to keep computation times reasonable. This leads to 69 (204) unique multiplets $\L$ and 120 (210) bin combinations $B$ for the 4PCF (5PCF), giving a total of $8,280$ ($42,840$) elements. Whilst this is certainly large, it is still much smaller than the number of galaxies in the survey, which is a practical upper-limit to the number of useful bins, if all are independent. As in \S\ref{subsec: edge-correction-in-practice}, we compute and average $\mathcal{N}_\L^B$ for all 32 data-minus-random splits as well as the normalization $\mathcal{R}_\L^B$ for a single chunk, before forming the edge-correction matrix $M_{\L\L''}$ of \eqref{eq: edge-correction-matrix} and solving for $\zeta_\L^B$ using \eqref{eq: edge-corrected-NPCF}. Using the CPU code on 16 Skylake cores, each split requires $\sim$\,$1\,$ ($4$) CPU-hours for the 4PCF (5PCF) computation, thus the entire statistic can be estimated in $\sim\,30$ ($120$) CPU-hours. The short computation time means that, in a real analysis, many mocks can be analyzed within the confines of standard computational allocations. Furthermore, estimating the 4PCF multiplets requires only twice as much computation time as estimating the 3PCF, for the same $\ell_\mathrm{max}$.

The edge-correction matrices, $M_{\L\L''}^B$ \eqref{eq: edge-correction-matrix}, corresponding to the above scenarios are shown in Fig.\,\ref{fig: coupling-matrices} \resub{for the 4PCF and 5PCF.}\footnote{\resub{For the 3PCF analogs of Figs.\,\ref{fig: coupling-matrices}\,\&\,\ref{fig: 4pcf-results}, see \citet{2015MNRAS.454.4142S}.}} The coupling matrix is a function of bin $B$; as expected, the coupling strength increases with the radial bin value since the CMASS survey window function is smooth except on the largest scales.  
For this reason we plot only the final radial bin. In both the 4PCF and 5PCF, the off-diagonal couplings between different angular momenta are small, at most $7\%$ ($5\%$) for the 4PCF (5PCF), consistent with that found for the 3PCF in \citet{2015MNRAS.454.4142S}. \resub{These may also be understood intuitively by way of simple edge-correction models, as in the former work.} Whilst Fig.\,\ref{fig: coupling-matrices} shows a matrix with non-trivial structure, almost all the largest couplings are found in closely separated multiplets $\L$ and $\L''$, \textit{i.e.} those with $|\ell_i-\ell_i'|\leq 1$. This motivates the earlier statement (\S\ref{subsec: edge-correction}) that, by measuring all multiplets up to some $\ell_\mathrm{max}$, we obtain accurate NPCF measurements up to $(\ell_\mathrm{max}-1)$.

\begin{figure}
    \centering
    \begin{subfigure}[t]{0.5\textwidth}
        \centering
        \includegraphics[width=\textwidth]{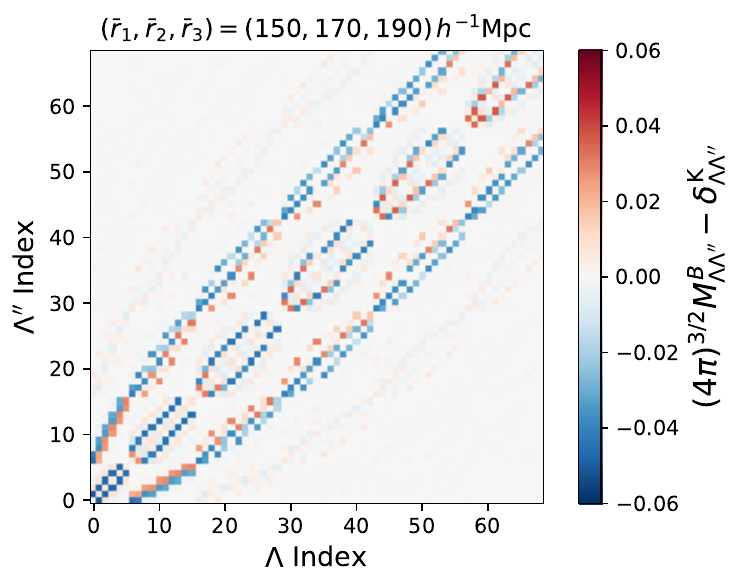}
        \caption{4PCF, $\ell_\mathrm{max} = 5$}
    \end{subfigure}%
    \begin{subfigure}[t]{0.5\textwidth}
        \centering
        \includegraphics[width=\textwidth]{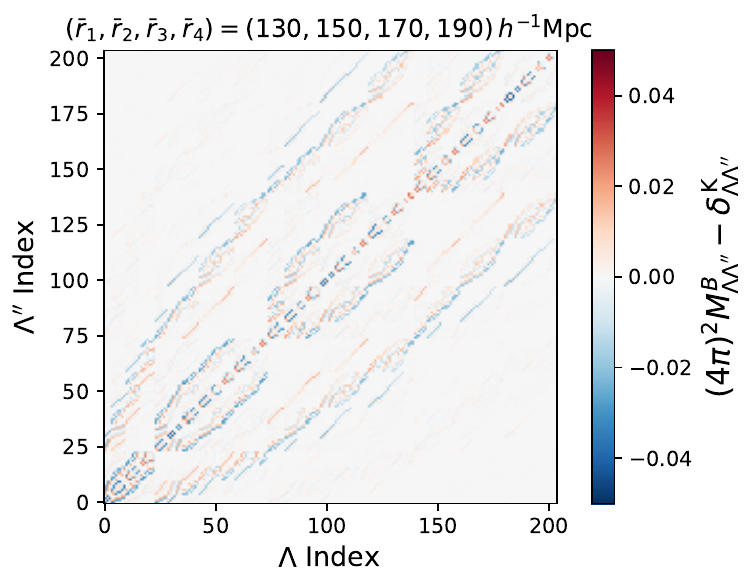}
        \caption{5PCF, $\ell_\mathrm{max} = 3$}
    \end{subfigure}%
    \caption{Edge-correction matrices $M_{\L\L''}^B$ for the 4PCF and 5PCF analyses of realistic \textsc{MultiDark-Patchy} simulations. As described in \S\ref{subsec: edge-correction}, these matrices give the couplings between multiplets due to the non-uniform survey geometry and are independent of the data itself. The horizontal and vertical axes give the indices of the $\L$ and the $\L''$ multiplets respectively, each of which specifies the angular momentum coefficients $\ell_1,\ell_2,\ldots,\ell_{N-1}$. These are arranged in ascending order with the last element of $\L$ varied first, and we omit any multiplets that do not satisfy the triangle conditions or are parity-odd, \textit{i.e.} the first few terms are $\L = \{0,0,0\}, \{0,1,1\}, \{0,2,2\}, \ldots \{1,0,1\}, \{1,1,0\} \ldots$ for the 4PCF. We show the matrices corresponding to the largest radial bin, since this contains the largest-amplitude correlations; the mean radius in each dimension is shown in the caption. The maximal off-diagonal coupling (in the $\L-\L''$ plane) ranges from $0.007$ to $0.06$ ($0.009$ to $0.05$) for the 4PCF (5PCF). The edge-correction matrices are normalized by $(4\pi)^{(N-1)/2}$ with the unit matrix subtracted, such that all values would be zero for a periodic box geometry. Note that the couplings between multiplets are small in all cases, and further, that, for any given $\ell_\mathrm{max}$, we do not find significant couplings between any multiplets with $|\ell_i-\ell_i'|>1$.}
    \label{fig: coupling-matrices}
\end{figure}

Finally, we come to the NPCF measurements themselves. Given that the statistic dimensionality is large, we here plot only a selection of $\L$ multiplets, but include all radial bins, collapsed into one-dimension. Fig.\,\ref{fig: 4pcf-results} shows the measured 4PCF multiplets for the mean of 10 \textsc{MultiDark-Patchy} mocks, and we immediately note that there is a strong signal in some components, particularly $\L = \{0,0,0\}$, $\{0,3,3\}$ and $\{2,0,2\}$, but a much weaker one in others, e.g., $\L = \{3,3,2\}$. It is important to stress that the trends in Fig.\,\ref{fig: 4pcf-results} do \textit{not} indicate a strong detection of non-Gaussianity; we must remember that the 4PCF contains both a disconnected contribution (from the product of two 2PCFs) and a connected component, as in \eqref{eq: 4pcf-discon-con}. The former is sourced by Gaussian statistics, and would be present even in the absence of non-Gaussianity; before binning, this information is fully degenerate with the 2PCF. For an isotropic Gaussian universe, the disconnected term contributes only to multiplets of the form $\L = \{0,\ell,\ell\}$ and permutations \citep{4pcf_boss}, thus the non-zero elements of $\L = \{1,1,2\}$ observed at small radius give evidence for violations of isotropy or Gaussianity; here they are primarily an indication of redshift-space distortions. 
This will be discussed in depth in \citet{4pcf_boss}, alongside a modified estimators that entirely remove the Gaussian contribution.

\begin{figure}
    \centering
    \includegraphics[width=0.95\textwidth]{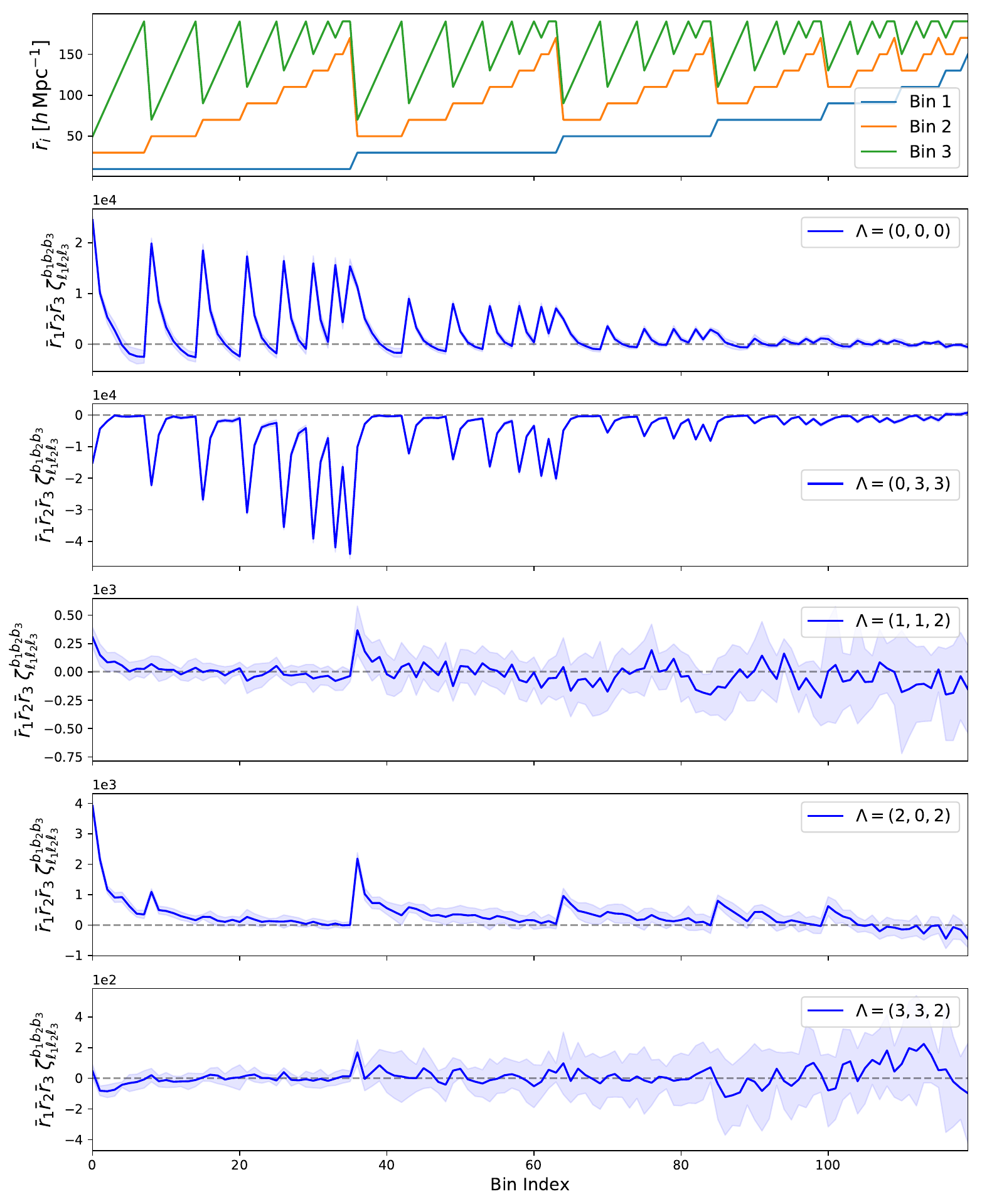}
    \caption{Edge-corrected 4PCF multiplets $\zeta_\L^B$ measured from 10 \textsc{MultiDark-Patchy} simulations for five choices of the multiplet index $\L = \{\ell_1,\ell_2,\ell_3\}$, \resub{setting the internal angles of the 4PCF tetrahedron}. We show the mean value (solid lines), and $1\sigma$ errors (shaded regions) in all cases. For each $\L$, there are $120$ choices of radial bin indices $B = \{b_1,b_2,b_3\}$; the top panel shows the linearly-averaged value of the radial bin, $\bar{r}$, corresponding to each. Succeeding panels show the individual multiplet estimates (normalizing by $\bar{r}_1\bar{r}_2\bar{r}_3$), with the value of $\L$ shown in the caption. Using $\ell_\mathrm{max}=5$, there are $69$ such multiplets, of which only five are shown. The `sawtooth' pattern of the 4PCF multiplets occurs due to the non-smooth variation in radial bins; a consequence of plotting a high-dimensional statistic in 1D. We note that the 4PCF contains both disconnected (Gaussian) and disconnected (non-Gaussian) contributions (cf.\,\S\ref{subsec: estimator-def}), thus much of this result is simply a detection of the 2PCF. All results are computed using 10 radial bins in the range $[0,200]\Mpch$ on a 16-core Intel Skylake machine. To compute and edge-correct the 4PCF in all multiplets $\L$ and bins $B$ for ten simulations, a total of $284$ CPU-hours was required.}
    \label{fig: 4pcf-results}
\end{figure}

Fig.\,\ref{fig: 5pcf-results} shows the analogous results for a selection of 5PCF multiplets. Again we note that some multiplets show a strong signal whilst others are noise-dominated; in contrast to the 4PCF, the disconnected term now involves the product of a 2PCF and 3PCF, thus Fig.\,\ref{fig: 5pcf-results} \textit{does} represent a robust detection of non-Gaussianity, albeit primarily in the 3PCF. A modified estimator may be derived to compute the connected-only 5PCF, as for the 4PCF. Analysis of the exact 5PCF shapes is beyond the scope of this work. However, we note a general trend of decreasing amplitude as the central bin values, $\bar{r}_i$ increase. Since late-time non-Gaussianity is usually concentrated on small scales, this is as expected.

\begin{figure}
    \centering
    \includegraphics[width=\textwidth]{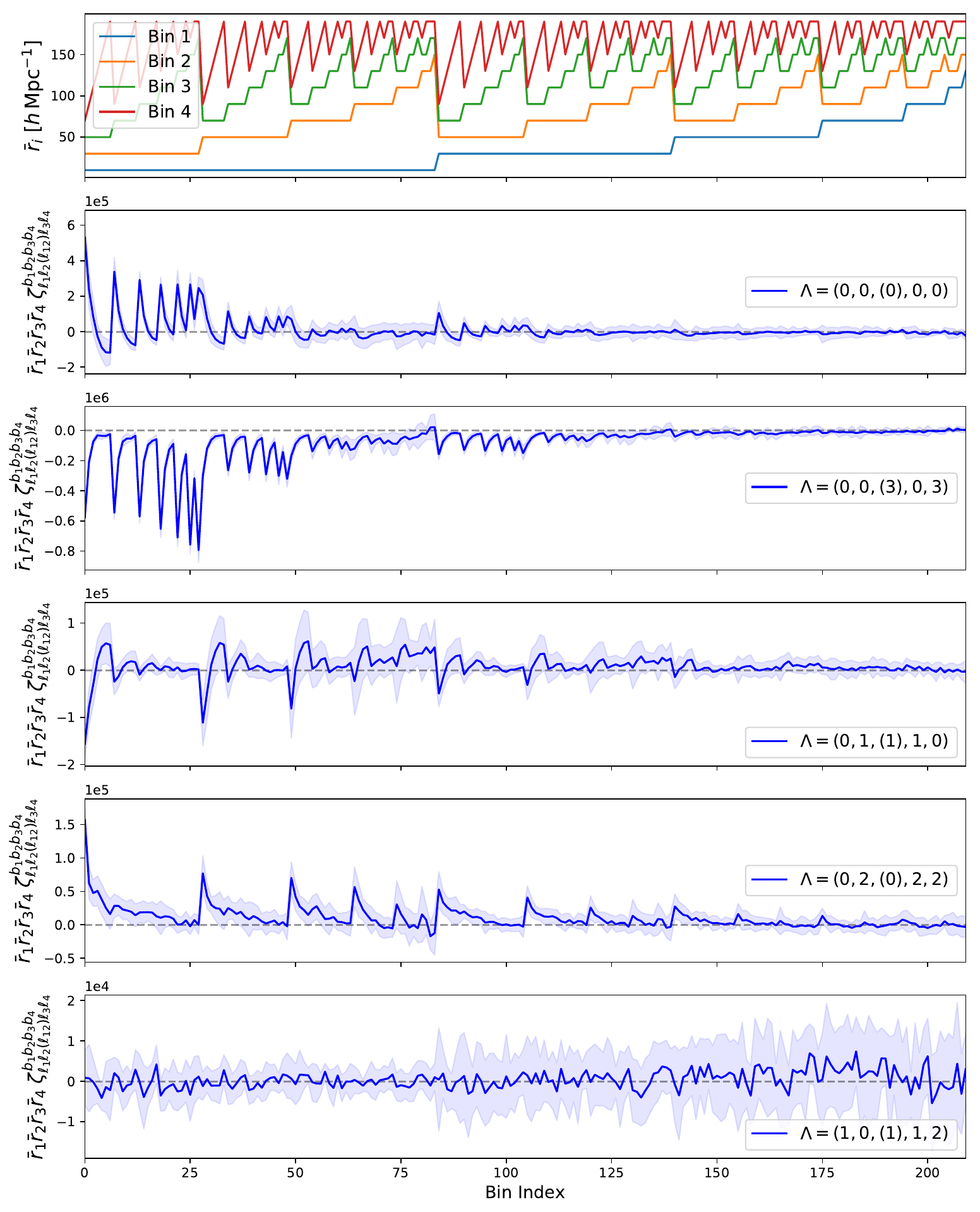}
    \caption{As Fig.\,\ref{fig: 4pcf-results} but for the 5PCF. As before, the first panel shows the bin-centers for each of the 210 radial bin combinations $B = \{b_1,b_2,b_3,b_4\}$, whilst the remaining plots give the measured multiplets $\zeta_\L^B$ with \resub{the internal angles specified by} $\L = \{\ell_1,\ell_2,(\ell_{12}),\ell_3,\ell_4\}$. Restricting to $\ell_\mathrm{max}=3$, there are $204$ such multiplets. As for the 4PCF, the signal is dominated by the disconnected contribution, here sourced by the product of a 2PCF and 3PCF. The total runtime for all 10 5PCF estimates was $1488$ CPU-hours.}
    \label{fig: 5pcf-results}
\end{figure}

\section{Summary}\label{sec: summary}

Correlation functions form the cornerstone of modern cosmology, and their efficient computation is a task of utmost importance for the analysis of current and future galaxy surveys. In this work, we have presented new algorithms for the estimation of the isotropic $N$-point correlation functions (NPCFs) for arbitrary order $N$. Making use of recently developed isotropic basis functions \citep{2020arXiv201014418C}, the estimator can be written in factorizable form and has complexity $\mathcal{O}(\Ng^2)$ for $\Ng$ particles. Furthermore, the output statistics are geometry-corrected and form a natural split into parity-even and parity-odd modes. The approach may be additionally generalized to homogeneous and isotropic spaces of arbitrary dimension, as discussed in our companion work \citep{npcf_generalized}.

We provide a public code, \textsc{encore},\footnote{\href{https://github.com/oliverphilcox/encore}{github.com/oliverphilcox/encore}} with which the isotropic 3PCF, 4PCF, 5PCF and 6PCF can be efficiently computed. The algorithm performs two major steps; a weighted pair-count over galaxies to compute $a_{\ell m}^b$ harmonic coefficients in radial bins $b$ for angular momentum coefficients $\ell, m$, and a summation over the $m$ indices of $(N-1)$ such harmonic coefficients to form the NPCF. Whilst the former scales as $\Ng^2$, the latter has $\mathcal{O}(\Ng)$ complexity, and is often found to be rate-limiting for $N>3$, particularly if the number of bins is large. In many cases, \textbf{our estimator scales linearly with the number of galaxies}. We have provided detailed discussion of the \textsc{encore} code, including its implementation in C\textsc{++} and \textsc{cuda}, with the latter allowing for GPU acceleration. Furthermore, results for the 4PCF and 5PCF of realistic mock catalogs have been presented; these are amongst the first of their kind. 
The efficiency of our approach allows them to be quickly estimated; each required $\sim$\,$30$ ($150$) CPU-hours in our test. Such methods will allow information to be extracted from the higher-point functions from upcoming surveys such as DESI and Euclid.

A number of avenues present themselves for future work. These include:
\begin{itemize}
    \item \textbf{Theory Model}: To place constraints on cosmology from the NPCFs, we require a robust theory model. Whilst straightforward for the Gaussian contributions, the non-Gaussian pieces are more difficult to treat and must be modeled perturbatively or with simulations.
    \item \textbf{Covariance}: Due to the increase in complexity of the basis functions as $N$ increases, the covariance matrices of the NPCF statistics are non-trivial, even in the isotropic and Gaussian limit. However, computing analytic covariances is of importance since the statistics are high-dimensional, prohibiting a full mock-based covariance. The Gaussian covariances will be presented in \citet{4pcf_cov}.
    \item \textbf{Connected Estimators}: For $N>3$, the NPCFs contain both \textit{disconnected} and \textit{connected} contributions (cf.\,\S\ref{subsec: estimator-def}). The former simply repeats information from the lower-point functions, and is not cosmologically relevant. Analyses of the higher-point functions must either model or subtract the disconnected pieces. A connected-only 4PCF estimator will be presented in \citet{4pcf_boss}.
    \item \textbf{Application to Data}: The techniques developed in this work allow computation of the edge-corrected NPCFs of survey data. Coupled with the connected-only estimator and Gaussian covariance matrices, this will allow the first estimation of the non-Gaussian higher-point functions from data. Furthermore, consideration of the parity-odd multiplets provides a test for parity-violation in the late Universe. Application to the BOSS galaxy survey will be considered in \citet{4pcf_boss}.
    \item \textbf{Gridded Estimator}: 
    Whilst the efficient estimator developed in \S\ref{sec: estimator} applies to both gridded and discrete density fields, only the latter is included in \textsc{encore}. Developing a gridded estimator (possible via FFTs) will allow fast analysis of high-density data and application to other contexts such as hydrodynamical simulations. This can be built upon the code of \citet{2018ApJ...862..119P}, which presented a gridded 3PCF estimator for analysis of the turbulent ISM. 
    \item \textbf{Anisotropic NPCFs}: Whilst we have focused only on the isotropic case in this work, anisotropic NPCFs also contain informmation. We expect a similar approach to allow estimation of these statistics, using a generalized angular momentum basis. This is briefly discussed in \citet{npcf_generalized}.
    \item \textbf{CMB Estimators}: A similar prescription to the above may be used to compute $N$-point statistics on the 2-sphere. This will allow \resub{$\mathcal{O}(N_\mathrm{pix}^2)$} estimation of CMB $N$-point functions from $N_\mathrm{pix}$ pixels, and is detailed in \citet{npcf_generalized}.
\end{itemize}

To close, we briefly reflect on the justification for measuring higher-order NPCFs, and possible alternatives. As is well-known, the process of density field reconstruction \citep[e.g.,][]{1999MNRAS.308..763M,2006MNRAS.365..939M,2007ApJ...664..675E,2009PhRvD..80l3501N} shifts information from the higher-point functions back to the 2PCF \citep{2015PhRvD..92l3522S} by reducing long-wavelength displacements. Given that it has become customary for survey analysis pipelines to implement BAO reconstruction \citep[e.g.,][]{2012MNRAS.427.2132P,2017MNRAS.464.3409B,2018MNRAS.477.1153V,2020JCAP...05..032P}, one must ask whether there remains any utility in the unreconstructed higher-point functions. As shown in \citet{2021MNRAS.505..628S}, the information on parameters affecting the \textit{evolution} of matter from its initial conditions, may be better measured by the combination of late-time NPCFs than the initial power spectrum (and hence reconstructed 2PCF). This occurs since these parameters appear in the forward model, rather than the initial conditions.
In essence, our argument is that the combination of the reconstructed 2PCF with unreconstructed NPCFs of higher order provides more information, particularly on parameters such as the dark energy density, than either one alone.

Ours is not the only way to proceed. An alternative approach involves forward-modeling the entire density field from its initial conditions to today \citep{2013MNRAS.432..894J,2019MNRAS.490.4237L,2017JCAP...12..009S,2018JCAP...07..043F,2019JCAP...10..035H}, including the possibility of field-level inference \citep{2019PhRvD.100d3514S,2021JCAP...05..059S,2020JCAP...11..008S,2020JCAP...04..042C,2021JCAP...04..032S}. Such approaches naturally capture information present both in the initial conditions and the forward model, and are a promising new way to perform inference. That said, analysis of this kind is computationally expensive due to the high dimensionality of the problem and necessity to run a large number of forward-model simulations to perform any parameter inference, though various tricks are used to expedite this. In contrast, the more classical approach of measuring the NPCFs allows fasts inference once the summary statistics have been computed and stored. Indeed, any new cosmological model can be tested without remeasuring any statistics, unlike in the forward-modeling framework. With the methods discussed in this work, including higher-order statistics in future analyses, such as those of DESI, Roman, and Euclid, has become an attainable goal.

\section*{Data Availability}
The data underlying this article will be shared on reasonable request to the corresponding author. The \textsc{encore} code is available at \href{https://github.com/oliverphilcox/encore}{github.com/oliverphilcox/encore} and the \textsc{MultiDark-Patchy} mocks can be downloaded from \href{https://data.sdss.org/sas/dr12/boss/lss/}{data.sdss.org/sas/dr12/boss/lss/}.



\section*{Acknowledgments}
\footnotesize
We are indebted to many people for insightful discussions, including, but not limited to: Giovanni Cabass, Simone Ferraro, J. Richard Gott, III, \resub{Colin Hill}, Chris Hirata, Benjamin Horowitz, Mikhail Ivanov, Alex Krolewski, Chung-Pei Ma, Marcel Schmittfull, Marko Simonovi\'c, David Spergel, \resub{James Sunseri}, Martin White, Matias Zaldarriaga, and all members of the Slepian research group. \resub{We additionally thank the referee for useful feedback, which helped improve the clarity of the manuscript.} 

OP acknowledges funding from the WFIRST program through NNG26PJ30C and NNN12AA01C \resub{and thanks the Simons Foundation for additional support}. ZS is grateful for hospitality at Laboratoire de Physique Nucl\'{e}aire et des Hautes Energies in Paris for part of the period of this work, as well as at Lawrence Berkeley National Laboratory. Support for this work was provided by the National Aeronautics and Space Administration through Einstein Postdoctoral Fellowship Award Number PF7-180167 issued by the Chandra X-ray Observatory Center, which is operated by the Smithsonian  Astrophysical Observatory for and on behalf of the National Aeronautics Space Administration under contract NAS8-03060. ZS also acknowledges support from a Chamberlain Fellowship at Lawrence Berkeley National Laboratory (held prior to the Einstein) and from the Berkeley Center for Cosmological Physics. RNC also thanks Laboratoire de Physique Nucl\'{e}aire et des Hautes Energies in Paris for their recurring  hospitality. RNC’s work is supported in part by the U.S. Department of Energy, Office of Science, Office of High Energy Physics under Contract No. DE-AC02-05CH11231. DJE was supported by DOE-SC0013718 and as a Simons Foundation Investigator.

The authors are pleased to acknowledge that the work reported on in this paper was substantially performed using the Princeton Research Computing resources at Princeton University which is consortium of groups led by the Princeton Institute for Computational Science and Engineering (PICSciE) and Office of Information Technology's Research Computing.



\appendix
\normalsize
\section{Generalized Gaunt Integrals}\label{appen: generalized-gaunt}
We give the explicit forms of the generalized Gaunt integrals \eqref{eq: generalized-gaunt-def} used to construct the edge-correction equations in \S\ref{subsec: edge-correction}. The forms at arbitrary order can be constructed following \citet[\S6]{2020arXiv201014418C}.

For $N=3$, $\L = \{\ell,\ell\}$, giving
\beq
    \mathcal{G}^{\L\L'\L''} =  \frac{1}{4\pi}\sqrt{(2\ell+1)(2\ell'+1)(2\ell''+1)}\tjo{\ell}{\ell'}{\ell''}^2,
\eeq
which is just a rescaled version of the integral of three Legendre polynomials, \textit{viz.} \eqref{eq: N=3-basis}. For $N=4$, $\L = \{\ell_1,\ell_2,\ell_3\}$, leading to
\beq
    \mathcal{G}^{\L\L'\L''} &=& \frac{1}{(4\pi)^{3/2}}\left[\prod_{i=1}^3\sqrt{(2\ell_i+1)(2\ell'_i+1)(2\ell_i''+1)}\tjo{\ell_i}{\ell'_i}{\ell''_i}\right]\times\begin{Bmatrix}\ell_1 &\ell_1' &\ell_1'' \\ \ell_2 & \ell_2' & \ell_2'' \\ \ell_3 & \ell_3' & \ell_3''\end{Bmatrix},
\eeq
where the $3\times3$ matrix is a Wigner 9-$j$ symbol \citep[e.g.,][\S34.6]{nist_dlmf}. Finally for $N=5$, $\L = \{\ell_1,\ell_2,(\ell_{12}),\ell_3,\ell_4\}$, and we obtain
\beq
    \mathcal{G}^{\L\L'\L''} &=& \frac{1}{(4\pi)^{2}}\sqrt{(2\ell_{12}+1)(2\ell'_{12}+1)(2\ell''_{12}+1)}\left[\prod_{i=1}^4\sqrt{(2\ell_i+1)(2\ell'_i+1)(2\ell_i''+1)}\tjo{\ell_i}{\ell'_i}{\ell''_i}\right]\\\nonumber
    &&\,\times\,\begin{Bmatrix}\ell_1 &\ell_2 &\ell_{12} \\ \ell_1' & \ell_{2}' & \ell_{12}' \\ \ell_1'' & \ell_{2}'' & \ell_{12}''\end{Bmatrix}\begin{Bmatrix}\ell_{12} &\ell_3 &\ell_4 \\ \ell_{12}' & \ell_{3}' & \ell_4' \\ \ell_{12}'' & \ell_{3}'' & \ell_{4}''\end{Bmatrix},
\eeq
now depending on two 9-$j$ symbols.

\section{Simplification of the $M$ Summations}\label{appen: spin-sum}
Below, we outline a simplification of the $M$ index summations necessary for the NPCF estimators. This reduces the computational requirements by a factor close to two. First, consider the summand of \eqref{eq: NPCF-mult-discrete-binned-estimator}, which must be computed at the location of each primary particle:
\beq\label{eq: summand-tmp}
    \mathcal{H}_\L^B(\vs_i) &\equiv& \sum_{m_1\ldots m_{N-1}}\mathcal{E}(\L)\;C^\L_{M}\prod_{i=j}^{N-1}a^{b_j}_{\ell_jm_j}(\vs_i).
\eeq
As noted in \S\ref{subsec: basis-def}, the coupling coefficients $C^\L_M$
impose $m_1+\ldots +m_{N-1} = 0$; this allows \eqref{eq: summand-tmp} to be written as a sum of three pieces:
\beq\label{eq: summand-tmp1}
    \mathcal{H}_\L^B(\vs_i) = \left(\sum_{m_1+\cdots +m_{n-2}<0}+\sum_{m_1+\cdots +m_{n-2}=0}+\sum_{m_1+\cdots +m_{n-2}>0}\right) \mathcal{E}(\L)\;C^\L_{M}\prod_{j=1}^{N-1}a^{b_j}_{\ell_jm_j}(\vs_i).
\eeq
Taking the final term and redefining $m_i\to -m_i$, we find
\beq\label{eq: summand-tmp2}
    \sum_{m_1+\cdots +m_{n-2}>0} \mathcal{E}(\L)\;C^\L_{M}\prod_{j=1}^{N-1} a_{\ell_jm_j}^{b_j}(\vs_i) &=& \sum_{m_1+\cdots +m_{n-2}<0} \mathcal{E}(\L)\;C^\L_{-M}\prod_{j=1}^{N-1} a_{\ell_j-m_j}^{b_j}(\vs_i)\\\nonumber
    &=& \sum_{m_1+\cdots +m_{n-2}<0} C^\L_{M}\prod_{j=1}^N(-1)^{m_j} a_{\ell_jm_j}^{b_j,*}(\vs_i),
\eeq
where we have used \eqref{eq: basis-coupling-M-inversion} and noted that $Y_{\ell -m}(\hr) = (-1)^mY_{\ell m}^*(\hr)$, thus $a_{\ell -m}^b(\vs_i) = (-1)^m a_{\ell m}^{b*}(\vs_i)$. The factor $\prod_{j=1}^{N-1}(-1)^{m_j} = 1$ by the summation rules. Combining \eqref{eq: summand-tmp2} with the first term in \eqref{eq: summand-tmp1} yields
\beq\label{eq: summand-tmp3}
    \left(\sum_{m_1+\cdots +m_{n-2}<0}+\sum_{m_1+\cdots +m_{n-2}>0}\right) \mathcal{E}(\L)\;C^\L_{M}\prod_{j=1}^N a_{\ell_jm_j}^{b_j}(\vs_i) &=& \sum_{m_1+\cdots+ m_{n-2}<0}\mathcal{E}(\L)\;C^\L_M\left(\prod_{j=1}^Na_{\ell_jm_j}^{b_j}(\vs_i)+\mathcal{E}(\L)\prod_{j=1}^Na_{\ell_jm_j}^{b_j,*}(\vs_i)\right)\\\nonumber
    &=& 2\sum_{m_1+\cdots +m_{n-2}<0}\mathcal{E}(\L)\;C^\L_M\times\mathbb{Q}_\L\left[\prod_{j=1}^Na_{\ell_jm_j}^{b_j}(r_j)\right],
\eeq
defining the operator
\beq
    \mathbb{Q}_\L[f] = \begin{cases}\mathrm{Re}[f] & \text{ if } \mathcal{E}(\L) = +1\\ i\,\mathrm{Im}[f] & \text{ if } \mathcal{E}(\L) = -1.\end{cases}
\eeq
Similarly, the second term in \eqref{eq: summand-tmp1} gives
\beq
    \sum_{m_1+\cdots +m_{n-2}=0} \mathcal{E}(\L)\;C^\L_{M}\prod_{j=1}^N a_{\ell_jm_j}^{b_j}(\vs_i) &=& \sum_{m_1+\cdots +m_{n-2}=0} C^\L_{M}\prod_{j=1}^N a_{\ell_jm_j}^{b_j,*}(\vs_i),
\eeq
relabelling $m_i\to-m_i$ and simplifying. Averaging the two yields the equivalent of \eqref{eq: summand-tmp3} but without the factor of $2$. In concert, we obtain
\beq
    \mathcal{H}_\L^B(\vs_i) = 2\sum_{m_1+\cdots +m_{n-2}\leq 0}S(m_{N-1})\; \mathcal{E}(\L)\;C^\L_{M}\; \mathbb{Q}_\L\left[\prod_{j=1}^Na_{\ell_jm_j}(\vs_i)\right],
\eeq
where $S(m) = 1/2$ if $m=0$ and unity else. This leads to a significant reduction in computation time and additionally makes manifest the notion that even-parity ($\mathcal{E}(\L)=+1$) multiplets are real and odd-parity ($\mathcal{E}(\L)=-1$) multiplets are imaginary.

\bibliographystyle{mnras}
\bibliography{bib,otherlib}

\bsp	
\label{lastpage}
\end{document}
